\documentclass[fleqn,usenatbib]{mnras}

\usepackage[T1]{fontenc}
\usepackage{ae,aecompl}

\usepackage{graphicx, float}	
\usepackage{amsmath}	
\usepackage{amssymb}	
\usepackage[hyphenbreaks]{breakurl}
\newcommand\maxij{MAXI~J1348--630}

\title[Broadband Properties of MAXI~J1348--630]{Broadband Spectral and Timing Properties of MAXI~J1348--630 using {\it \textbf{AstroSat}} and {\it \textbf{NICER}} Observations}

\author[Jithesh et al.]{
V. Jithesh$^{1}$\thanks{E-mail: vjithesh@iucaa.in},
Ranjeev Misra$^{1}$, Bari Maqbool$^{1,2}$ and Gitika Mall$^3$ \\
$^{1}$Inter-University Centre for Astronomy and Astrophysics (IUCAA), PB No.4, Ganeshkhind, Pune-411007, India\\
$^{2}$Department of Physics, Islamic University of Science and Technology, Awantipora, Kashmir-192122, India\\ 
$^{3}$Center for Field Theory and Particle Physics and Department of Physics, Fudan University, 200438 Shanghai, China\\ 
}

\date{Accepted XXX. Received YYY; in original form ZZZ}

\pubyear{2021}

\begin{document}
\label{firstpage}
\pagerange{\pageref{firstpage}--\pageref{lastpage}}
\maketitle

\begin{abstract}

We present broadband X-ray spectral-timing analysis of the new Galactic X-ray transient MAXI~J1348--630 using five simultaneous {\it AstroSat} and {\it NICER} observations. Spectral analysis using {\it AstroSat} data identify the source to be in the soft state for the first three observations and in a faint and bright hard state for the next two. Quasi-periodic oscillations at $\sim 0.9$ and $\sim 6.9$\,Hz, belonging to the type-C and type-A class are detected. In the soft state, the power density spectra are substantially lower (by a factor $> 5$) for the {\it NICER} (0.5--12 keV) band compared to the {\it AstroSat}/LAXPC (3--80 keV) one, confirming that the disk is significantly less variable than the Comptonization component. For the first time, energy-dependent fractional rms and time lag in the 0.5--80 keV energy band was measured at different Fourier frequencies, using the bright hard state observation. Hard time lag is detected for the bright hard state, while the faint one shows evidence for soft lag. A single-zone propagation model fits the LAXPC results in the energy band 3--80 keV with parameters similar to those obtained for Cygnus X--1 and MAXI J1820+070. Extending the model to lower energies, reveals qualitative similarities but having quantitative differences with the {\it NICER} results. These discrepancies could be because the {\it NICER} and {\it AstroSat} data are not strictly simultaneous and because the simple propagation model does not take into account disk emission. The results highlight the need for more joint coordinated observations of such systems by {\it NICER} and {\it AstroSat}.
\end{abstract}




\section{Introduction}
 
Black hole transients (BHTs) are usually discovered when they exhibit outbursts, which are characterized by distinct spectral and temporal states \citep[see ][for reviews]{Rem06, Bel11}. Based on the spectro-timing properties, four main states have been identified in BHTs: the hard state (HS), the soft state (SS), the hard and soft intermediate states (HIMS and SIMS). In the HS, the source is characterized by a hard spectrum with a typical photon index $\Gamma \sim 1.7$. The power density spectra (PDS) are dominated by strong broad-limited noise with typical root mean square (rms) values of $\sim 30\%$ \citep{Bel05} and occasionally exhibits type-C low-frequency quasi-periodic oscillations (LFQPOs) along with a sub-harmonic or second harmonic in the PDS. Type-C QPOs are characterized by a narrow peak with centroid frequency ranging from few mHz to $\sim 30$\,Hz \citep{Rem02, Cas05}. The SS spectra are dominated by a thermal disk component and the variability amplitude reduces to a few percent. Weak QPOs with a frequency range of 6--8 Hz are sometimes detected in the SS, which belong to the so-called type-A category \citep{Wij99, Cas04, Mot11, Mot16}. In the HIMS and SIMS, the energy spectrum is a combination of soft and hard components, while the PDS contains type-C LFQPOs in HIMS and type-A and B in the SIMS.

From 1996 to 2012, the {\it Rossi X-ray Timing Explorer} ({\it RXTE}) was the workhorse in the field of rapid time variability of X-ray binaries. Now the Large Area X-ray Proportional Counter (LAXPC) onboard {\it AstroSat} has replaced {\it RXTE} and is contributing to the rapid time variability studies in the hard energy band. For example, {\it AstroSat} data has been used to study the spectral-timing properties of several black hole X-ray binaries, including Cygnus X--1 \citep{Mis17, Maq19}, Cygnus X--3 \citep{Pah17}, MAXI J1535--571 \citep{Bha19, Sre19}, Swift J1658.2--4242 \citep{Jit19}, GRS 1915+105 \citep{Raw19, Bel19, Mis20, Sre20}, MAXI J1820+070 \citep{Mud20} and 4U 1630--472 \citep{Bab20}. The soft X-ray ($< 4$ keV) rapid timing properties of black hole X-ray binaries (BHXRBs) were largely unknown, which is now being explored using the X-ray Timing Instrument (XTI) onboard the Neutron star Interior Composition Explorer ({\it NICER}). {\it NICER} observations have provided unprecedented soft X-ray timing characteristics of several black hole systems, MAXI J1535--571 \citep{Sti18, Ste18} MAXI J1820+070 \citep{Kar19, Sti20, Hom20} and MAXI J1348--630 \citep{Bel20, Zha20}. However, broadband (0.3--30 keV) fast timing properties of BHXRBs has been relatively less studied. There has been one such attempt to the understand the broadband spectral-timing behaviour of the transient BHXRB Swift J1658.2--4242 using simultaneous {\it Insight-HXMT}, {\it NICER} and {\it AstroSat} observations \citep{Xia19}, where a QPO at $\sim 1.5$\,Hz was detected in all three satellites. The study further emphasizes the need for simultaneous broadband observations from different satellites to understand the fast timing properties of BHXRBs. 

\maxij{} is a new X-ray transient source discovered by the {\it MAXI}/GSC instrument on 2019 January 26 \citep{Yat19}. {\it Swift} XRT observation localized the source position with the reported position being R.A. = 13:48:12.73, Decl. = -63:16:26.8 (equinox J2000.0) with an uncertainty of $\sim 1.7$ arcsec \citep[90\% confidence level;][]{Ken19}. The source was observed by all major X-ray observatories like {\it INTEGRAL}, {\it NICER} and {\it Insight-HXMT} \citep{Lep19, San19, Che19}. An optical counterpart has been identified for the source with iTelescope.Net T31 instrument in Siding Spring, Australia \citep{Den19}. Radio observation with the {\it Australia Compact Telescope Array} ({\it ATCA}) detected a radio source consistent with the X-ray position and combined radio and X-ray properties suggest that the source is a BHXRB \citep{Rus19}. The detailed X-ray spectral study using the first half-year MAXI/GSC monitoring observations suggests that \maxij{} may host a relatively massive spinning black hole with a mass of $\sim 16~\rm M_{\sun}$ \citep{Tom20}. Using the broadband energy spectrum from {\it Swift} XRT, BAT and {\it MAXI}/GSC observations, \citet{Jan20} estimated the black hole mass as $\sim 9~\rm M_{\sun}$ based on the two-component advective flow model.

\citet{Bel20} studied a set of {\it NICER} observations of \maxij{} during the brightest part of the outburst and detected a strong type-B QPO at $\sim 4.5$ Hz. The fractional rms at the QPO frequency increased from $< 1\%$ to $> 10 \%$ in the {\it NICER} energy band and a hard lag at the QPO frequency was detected. The energy spectrum was fitted by a thin disk plus a steep hard power law component along with an emission line in the 6--7 keV band. The spectral analysis suggested that source was in the SIMS. Recently, \citet{Zha20} performed a detailed analysis of the outburst evolution and timing properties of the source using {\it NICER} observations. During the outburst the source evolved from the HS into the SS through HIMS and SIMS and made a transition back to the HS during the outburst decay. In addition to the main outburst, the source exhibited two reflares with much lower peak intensity compared to the main outburst and remained in the hard spectral state. They also detected type-A, type-B and type-C QPOs at different phases of the outburst.   
  
In this paper, we study the broadband X-ray spectral and timing characteristics of \maxij{} using simultaneous {\it AstroSat} and {\it NICER} observations. Section \ref{sec:obs} describes the observations used in this work and the data reduction techniques. The broadband spectral and timing analysis are presented in \S \ref{sec:spectral} and \S \ref{sec:timing}, respectively. We modelled the energy-dependent timing properties of the source using stochastic propagation model, which is described in \S \ref{sec:modelling}. The main results are summarised and discussed in \S \ref{sec:discu}. 

\section{Observations and Data Reduction}
\label{sec:obs}

\begin{figure}
\begin{center}

\includegraphics[width=8.5cm,angle=0]{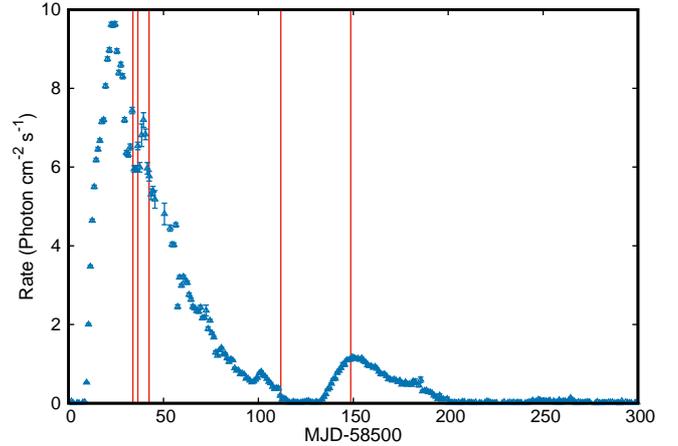}
\caption{The one-day binned 2--20 keV {\it MAXI} light curve of \maxij{} over the period 2019 January 17 to November 13. The vertical red lines represent {\it AstroSat} observations considered in this work.}
\label{maxi-lc}
\end{center}
\end{figure}

\subsection{{\textbf{\it AstroSat}}}
We used five publicly available {\it AstroSat} \citep{Sin14, Agr17} Target of Opportunity (ToO) observations of \maxij{}. We marked the observations used in this work on the {\it MAXI} light curve, which is shown in Figure \ref{maxi-lc} and their details are listed in Table \ref{obslog}.


\begin{table*}
\centering
\setlength{\tabcolsep}{16.0pt}
	\caption{Observation Log. (1) Observation data (AS and N represent {\it AstroSat} and {\it NICER}, respectively); (2) observation ID; (3) date of observation; (4) exposure time (L and S represent LAXPC and SXT, respectively); (5) SXT count rate from a circular region of radius 16 arcmin; (6) inner and outer radius of annulus used for extraction of SXT spectrum. $^a$The SXT data is not piled-up in AS4 observation. Thus, a circular region of radius 16 arcmin is used for event extraction.}
 	\begin{tabular}{@{}cccccc@{}}
	\hline
	\hline
Data & ObsID & Date & Exposure & SXT Count Rate & Radius \\
 & & & (ks) & (c/s) & (arcmin) \\
\hline
AS1 & T03\_083T01\_9000002722 & 2019 February 19--20 & 5.5(L)/1.9(S) & 879.4 & 8 \& 16 \\ 
N1  & 1200530118 & 2019 February 19 & 5.0 & \\

AS2 & T03\_083T01\_9000002728 & 2019 February 22 & 20.2(L)/11.1(S) & 844 & 8 \& 16 \\
N2  & 1200530121 & 2019 February 22 & 2.5 & \\

AS3 & T03\_083T01\_9000002742 & 2019 February 28 & 23.2(L)/12.2(S) & 765.7 & 6 \& 16 \\
N3  & 1200530127 & 2019 February 28 & 2.8 & \\ 

AS4 & T03\_112T01\_9000002896 & 2019 May 8--9    & 13.8(L)/6.8(S) & 13.7 & 16$^a$ \\
N4  & 2200530133 & 2019 May 9 & 1.9 & \\ 

AS5 & T03\_120T01\_9000002990 & 2019 June 14--15 & 35.0(L)/14.9(S) & 68.7 & 2 \& 16 \\
N5  & 2200530154 & 2019 June 14 & 1.8 & \\ 
N6  & 2200530155 & 2019 June 15 & 1.6 & \\ 

\hline
\end{tabular} 
\label{obslog}
\end{table*}

\subsubsection{Large Area X-ray Proportional Counter}
LAXPC consists of three proportional counters ({\tt LAXPC10, LAXPC20} and {\tt LAXPC30}) operating in the energy range of 3--80 keV with a temporal resolution of $10\,\rm \mu s$ \citep{Yad16, Yad16a, Ant17, Agr17a}. We processed the Event Analysis (EA) mode data from these observations using LAXPC software\footnote{\url{http://astrosat-ssc.iucaa.in/?q=laxpcData}} (LaxpcSoft; version as of 2020 August 04). We applied the barycenter correction to the LAXPC level 2 data using the {\tt as1bary} tool. The standard tools available in LaxpcSoft \footnote{\url{http://www.tifr.res.in/~astrosat\_laxpc/LaxpcSoft.html}} were used to extract the light curves and energy spectra. The {\tt LAXPC10} detector was operating at low gain and the {\tt LAXPC30} detector was switched off on 2018 March 8 due to the gas leakage. Thus, we used {\tt LAXPC20} detector for our analysis. We modelled the {\tt LAXPC20} spectrum in the 5--40 keV energy band.

\subsubsection{Soft X-ray Telescope}
Soft X-ray Telescope (SXT) is a focussing telescope \citep{Sin16, Sin17} and all the observations are taken in Photon Counting (PC) mode. SXT has a large time resolution of 2.38\,s in PC mode compared to LAXPC. The Level-1 data were processed using the SXT pipeline software\footnote{\url{http://www.tifr.res.in/~astrosat\_sxt/sxtpipeline.html}} (version: AS1SXTLevel2-1.4b) to obtain the cleaned Level-2 event files for each orbit. We merged different orbits data using the SXT event merger tool\footnote{\url{http://www.tifr.res.in/~astrosat\_sxt/dataanalysis.html}\label{link2}} (Julia based module) and obtained an exposure corrected, merged cleaned event file. The source was piled-up in all observations except AS4 and we removed the pile-up by extracting the source events from annulus region. Different inner radii are used to mitigate the pile-up from SXT data and these radii are given in Table \ref{obslog}. For AS4 observation, we extracted the spectrum from a circular region of radius 16 arcmin. The blank sky SXT spectrum and the redistribution matrix file (sxt\_pc\_mat\_g0to12.rmf) provided by the instrument team were used as the background spectrum and RMF, respectively. The {\tt sxtARFModule} tool$^2$ were used to generate the SXT off-axis auxiliary response files (ARF) using on-axis ARF (sxt\_pc\_excl00\_v04\_20190608.arf), provided by the SXT instrument team. While fitting the SXT spectrum, we modified the gain of the response file by the {\tt gain} command, where the slope is fixed at unity and offset is a free parameter. We used the SXT spectrum in the 0.8--7 keV energy range. 
\subsection{{\it NICER}}
{\it NICER} \citep{Gen12} is a payload onboard International Space Station (ISS). The X-ray Timing Instrument (XTI) of {\it NICER} comprises of 56 X-ray optics with silicon detectors operating in the 0.2--12 keV energy band  \citep{Gen16}. Currently, 52 detectors are active. {\it NICER} monitored \maxij{} from 2019 January 26 immediately after the detection by {\it MAXI}/GSC. The source was observed by {\it NICER} for more than 300 ks. In this work, we used those {\it NICER} observations, which are simultaneous to {\it AstroSat}. Hence, we used six {\it NICER} observations and their details are given in Table \ref{obslog}. Among them, the observation N2 had telemetry saturation and as a result, the event data was fragmented into very short duration data segments\footnote{\url{https://heasarc.gsfc.nasa.gov/docs/nicer/data\_analysis/nicer\_analysis\_tips.html}}. In this case, we have omitted the MPU0, 3, 4 and 5 and continued our analysis. We have processed the data using {\sc heasoft} version 6.26.1, {\it NICER} software version 2019-06-19\_V006a and {\it NICER} CALDB version of 20200202 by applying standard filter criteria. We have further excluded detectors \#14 and \#34 from all observations, which show increased electronic noise occasionally. We examined for the presence of high-energy background flares by extracting the light curve in the 12--15 keV energy band \citep[see][]{Bul18}. No high background intervals were found in the {\it NICER} observations. We use the {\sc ftool barycorr} to apply the barycenter correction for each {\it NICER} observation. Since the source did not show any spectral and intensity variations in the N5 and N6 observations, we combined them for the analysis.    

\begin{figure*}
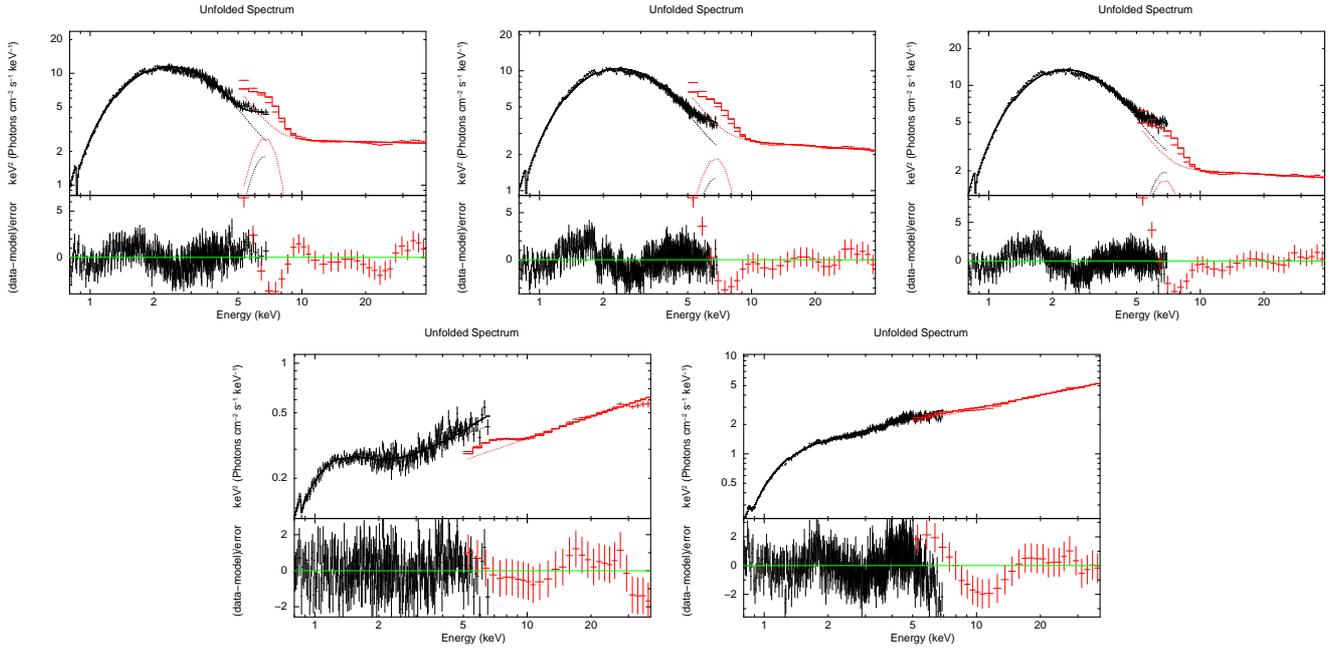


 \includegraphics[width=4.25cm,angle=-90]{f2a.ps}
 \includegraphics[width=4.25cm,angle=-90]{f2b.ps}
 \includegraphics[width=4.25cm,angle=-90]{f2c.ps}

 \includegraphics[width=4.25cm,angle=-90]{f2d.ps}
 \includegraphics[width=4.25cm,angle=-90]{f2e.ps}

\caption{The 0.8--40 keV broadband X-ray spectrum of \maxij{} from AS1 (top left), AS2 (top middle), AS3 (top right), AS4 (bottom left) and AS5 (bottom right) observations. The black and red data points represent the SXT and {\tt LAXPC20} data, respectively. Spectra are fitted with {\tt tbabs*(simpl * diskbb + gaussian)} model.}

\label{5spec}
\end{figure*}

\begin{table*}
\setlength{\tabcolsep}{3.0pt}
	\caption{Broadband X-ray Spectral Parameters for \maxij{}. (1) Observation; (2) segment used; (3) neutral hydrogen column density in units of $10^{22}~\rm cm^{-2}$ from {\tt tbabs} model; (4) photon power law index; (5) scattered fraction; (6) inner disk temperature in keV  (7) line energy in keV; (8) line width in keV; (9) {\tt gaussian} normalization; (10)-(11) the unabsorbed disk flux and total flux in units of $10^{-9}\rm~erg~cm^{-2}~s^{-1}$ in the 0.8--40 keV band derived using {\tt cflux}; (12) Ratio of the disk flux to the total flux; (13) $\chi^2$ statistics and degrees of freedom. In AS2, L and H represent the low and high intensity level segments.}
 	\begin{tabular}{@{}ccccccccccccc@{}}
	\hline
	\hline
Obs & Seg & $N_{\rm H}$ & $\Gamma$ & $f$ & $\rm kT_{in}$ & $E_{line}$ & $\sigma$ &$\rm N_{gauss}$ & $\rm F_{disk}$ & $\rm F_{Total}$ & $\rm Ratio$ & $\rm \chi^2/ d.o.f$ \\

\hline

AS1 & & $0.54^{+0.01}_{-0.01}$ & $2.05^{+0.03}_{-0.03}$ & $0.065^{+0.004}_{-0.004}$ & $0.799^{+0.007}_{-0.007}$ & 6.4 (f) & $1.00^{+0.09}_{-0.09}$ & $0.107^{+0.012}_{-0.011}$ & $32.03^{+0.21}_{-0.21}$ & $39.89^{+0.34}_{-0.33}$ & 0.80 & $570.0/395$ \\

AS2 & & $0.51^{+0.01}_{-0.01}$ & $2.12^{+0.03}_{-0.03}$ & $0.080^{+0.005}_{-0.005}$ & $0.790^{+0.005}_{-0.006}$ & 6.4 (f) & $1.14^{+0.09}_{-0.09}$ & $0.084^{+0.008}_{-0.008}$ & $28.64^{+0.16}_{-0.16}$ & $36.33^{+0.24}_{-0.23}$ & 0.79 & $743.1/514$ \\

AS2 & L & $0.50^{+0.01}_{-0.01}$ & $2.03^{+0.03}_{-0.03}$ & $0.054^{+0.004}_{-0.003}$ & $0.781^{+0.005}_{-0.005}$ & 6.4 (f) & $1.04^{+0.10}_{-0.10}$ & $0.064^{+0.007}_{-0.007}$ & $28.08^{+0.16}_{-0.16}$ & $33.86^{+0.24}_{-0.24}$ & 0.83 & $688.7/447$ \\

AS2 & H & $0.51^{+0.01}_{-0.01}$ & $2.19^{+0.03}_{-0.03}$ & $0.100^{+0.007}_{-0.006}$ & $0.793^{+0.007}_{-0.007}$ & 6.4 (f) & $1.10^{+0.09}_{-0.09}$ & $0.094^{+0.010}_{-0.010}$ & $29.22^{+0.19}_{-0.19}$ & $38.32^{+0.27}_{-0.27}$ & 0.76 & $676.5/480$ \\

AS3 & & $0.51^{+0.01}_{-0.01}$ & $2.09^{+0.03}_{-0.03}$ & $0.078^{+0.005}_{-0.005}$ & $0.769^{+0.005}_{-0.005}$ & 6.4 (f) & $1.14^{+0.07}_{-0.07}$ & $0.128^{+0.011}_{-0.010}$ & $36.98^{+0.21}_{-0.21}$ & $47.45^{+0.31}_{-0.31}$ & 0.78 & $814.9/538$ \\

AS4 & & $0.39^{+0.03}_{-0.03}$ & $1.57^{+0.02}_{-0.02}$ & $0.198^{+0.027}_{-0.025}$ & $0.277^{+0.015}_{-0.013}$ & 6.4 (f) & $1.37^{+0.33}_{-0.31}$ & $0.005^{+0.002}_{-0.002}$ & $0.46^{+0.03}_{-0.03}$ & $3.17^{+0.07}_{-0.07}$ & 0.15 & $269.6/300$ \\

AS5 & & $0.51^{+0.01}_{-0.01}$ & $1.55^{+0.01}_{-0.01}$ & $0.482^{+0.022}_{-0.031}$ & $0.330^{+0.016}_{-0.009}$ & 6.4 (f) & $1.97^{+0.45}_{-0.33}$ & $0.017^{+0.007}_{-0.005}$ & $1.66^{+0.08}_{-0.08}$ & $18.12^{+0.27}_{-0.19}$ & 0.09 & $551.1/492$ \\

\hline
\end{tabular} 
\label{spec_params}
\end{table*}

\begin{figure*}

 \includegraphics[width=5.8cm,angle=0]{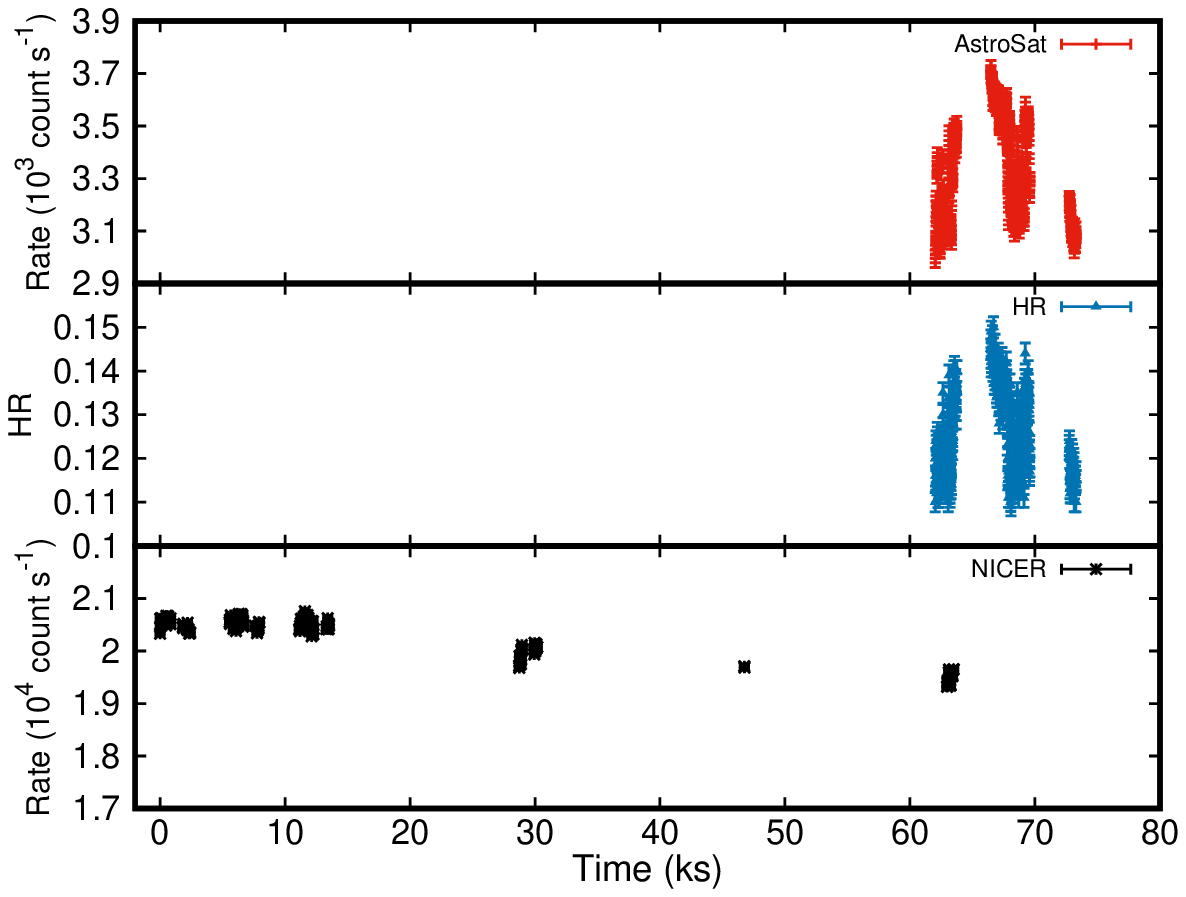}
 \includegraphics[width=5.8cm,angle=0]{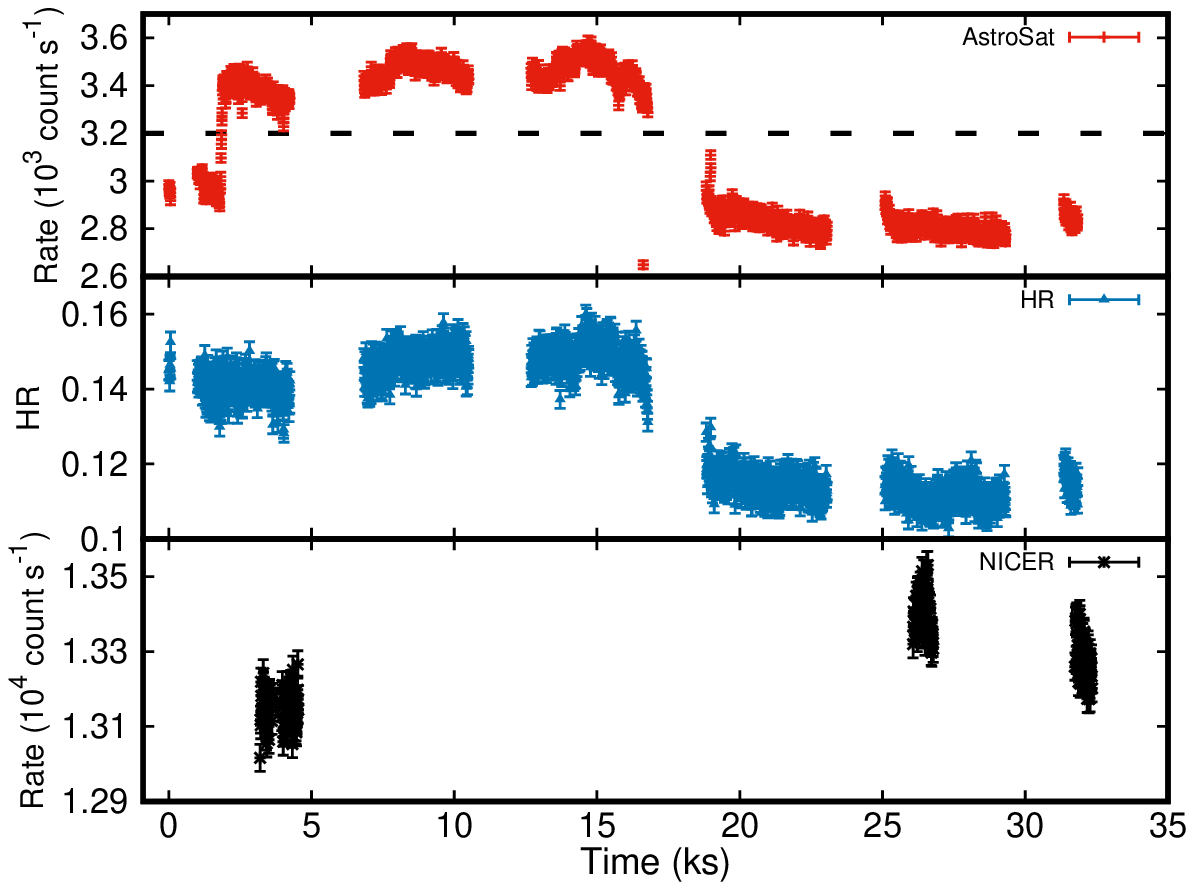}
 \includegraphics[width=5.8cm,angle=0]{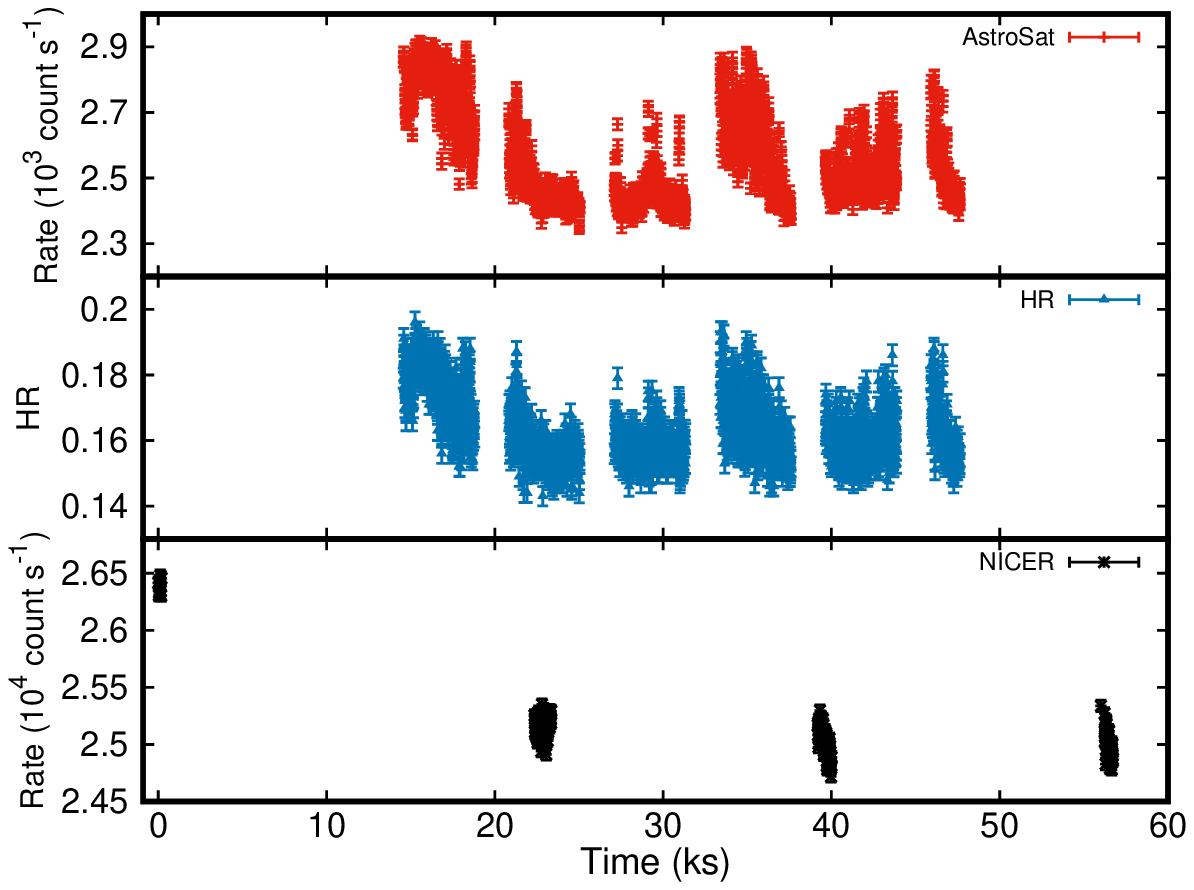}

\caption{The light curves and hardness ratio from simultaneous {\it AstroSat} and {\it NICER} observations in the soft state: AS1 \& N1 (left), AS2 \& N2 (middle) and AS3 \& N3 (right). In each plot, the top, middle and bottom panels represent the 3--80 keV light curve from {\tt LAXPC20} binned with 10-s, hardness ratio from {\it AstroSat} data and 10-s binned light curve in the 0.5-12 keV energy band from {\it NICER} observations, respectively. The dashed horizontal line in AS2 observation represents the two flux levels. Hardness ratio is defined as the ratio of the count rate in 7--16 keV  and the 3--7 keV bands.}
\label{lc_soft}
\end{figure*}

\begin{figure}

 \includegraphics[width=8.20cm,angle=0]{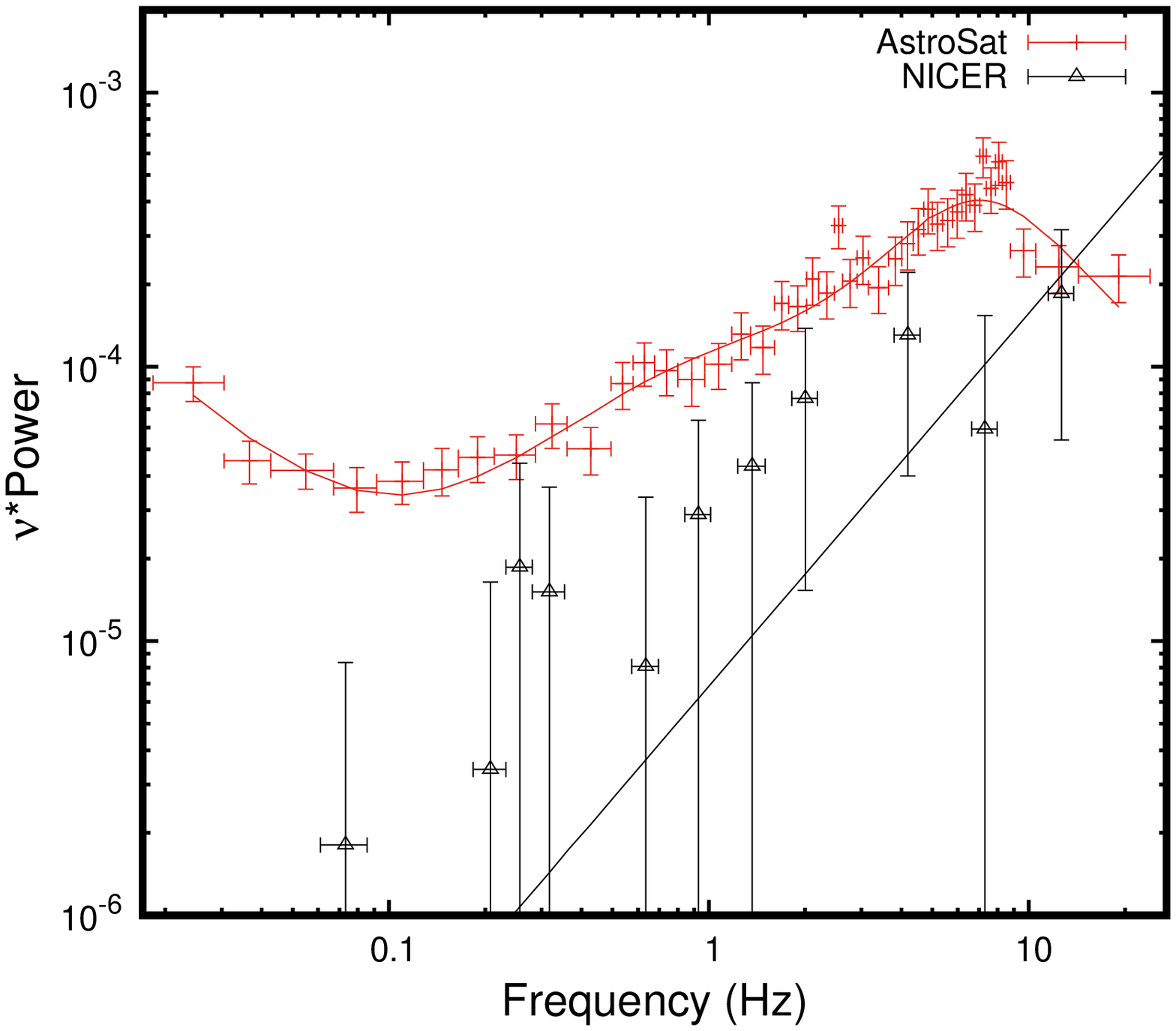}
 \includegraphics[width=8.20cm,angle=0]{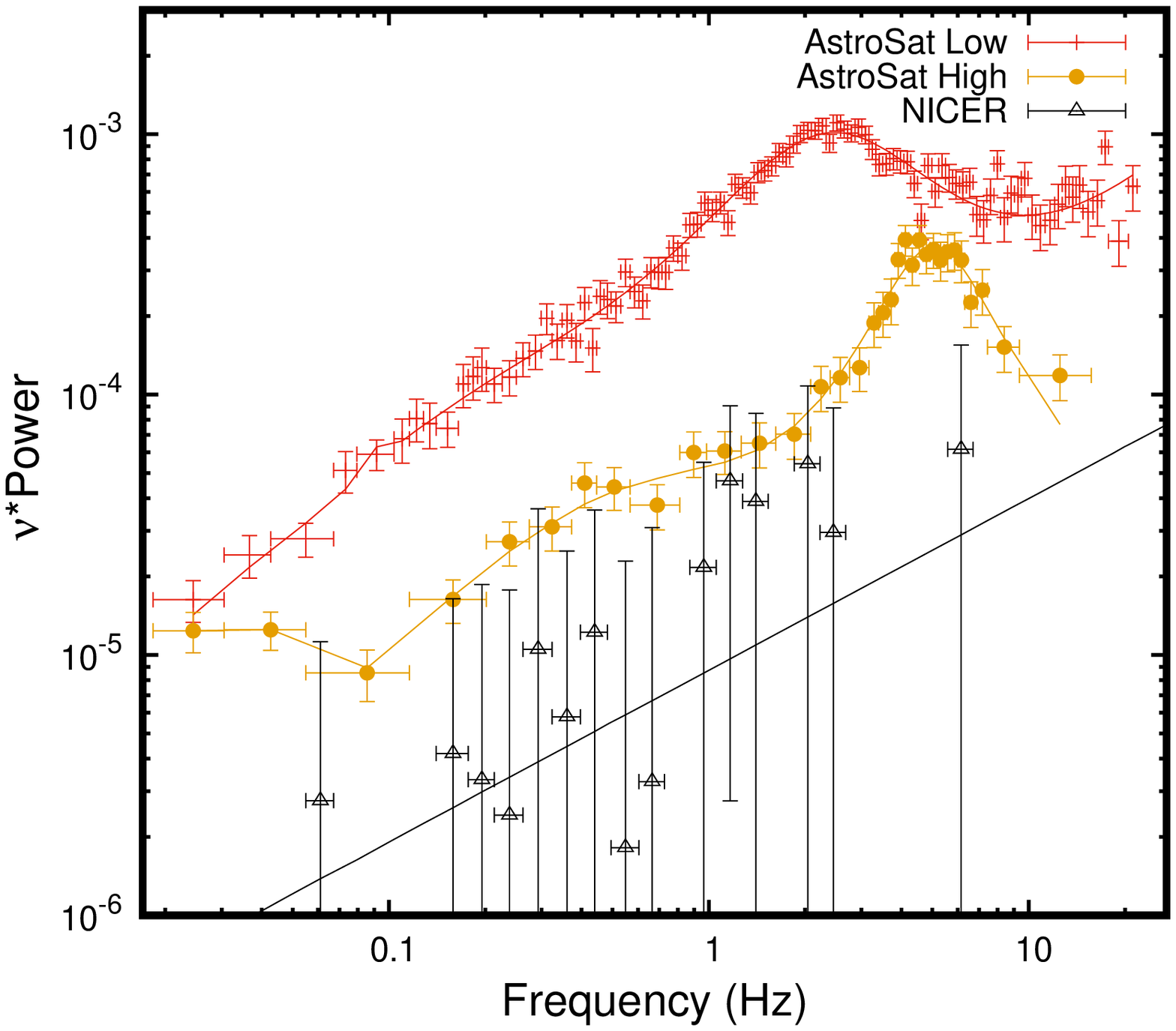}
 \includegraphics[width=8.20cm,angle=0]{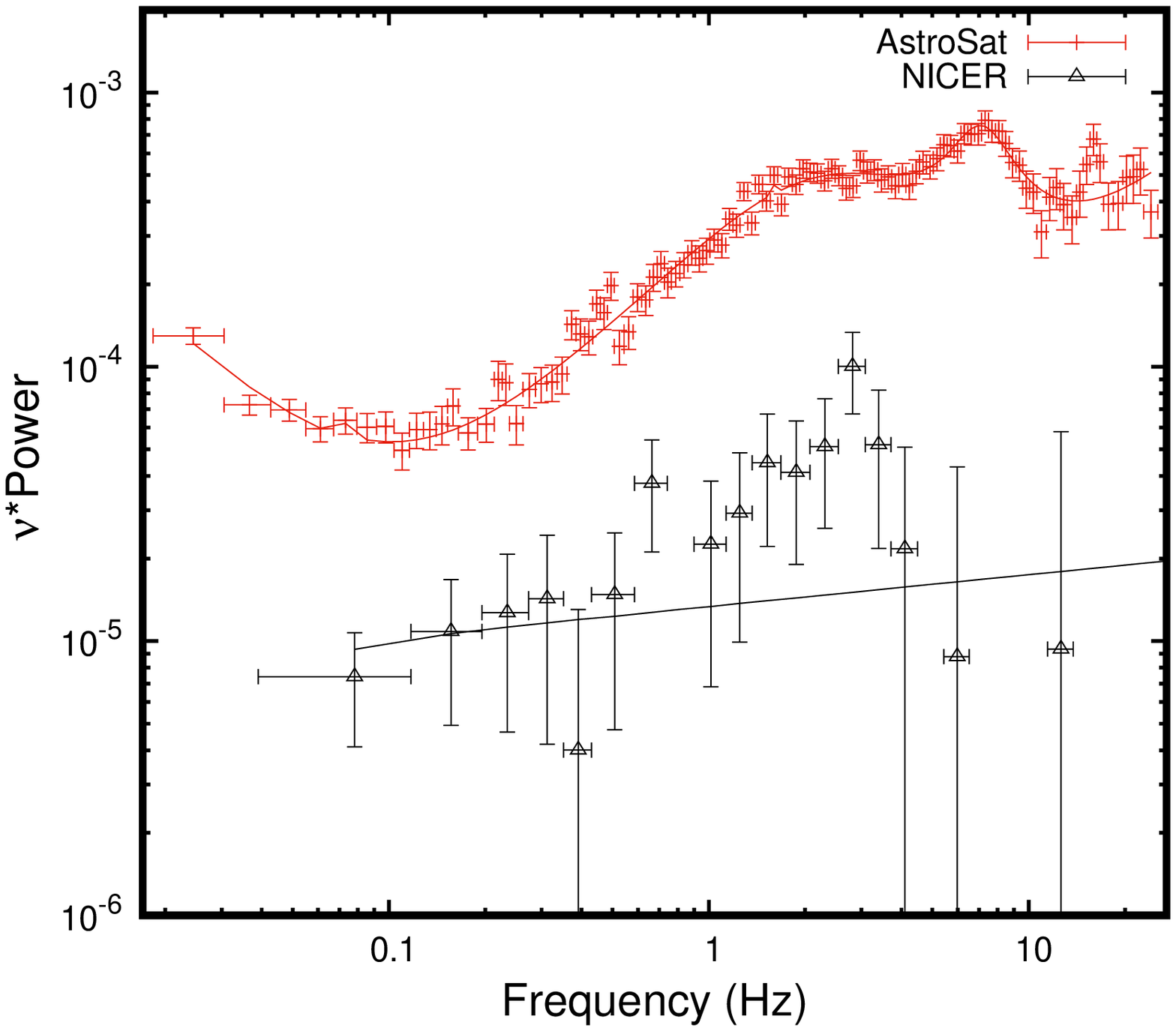}

\caption{The power density spectrum of \maxij{} from {\it AstroSat}/LAXPC (red plus points) and {\it NICER} (black open triangles) observations in the soft state. The observations are AS1 \& N1 (top), AS2 \& N2 (middle), AS3 \& N3 (bottom). In AS2, red plus points and yellow filled circles represent the PDS from the low and high intensity levels in LAXPC light curve (see middle panel of Figure \ref{lc_soft}). The PDS from AS1 and AS2 are fitted with three Lorentzians, while the AS3 PDS is fitted with four Lorentzians. The {\it NICER} PDS are fitted with a power law function.}
\label{pds_first3}
\end{figure}

\begin{figure*}

\includegraphics[width=5.80cm,angle=0]{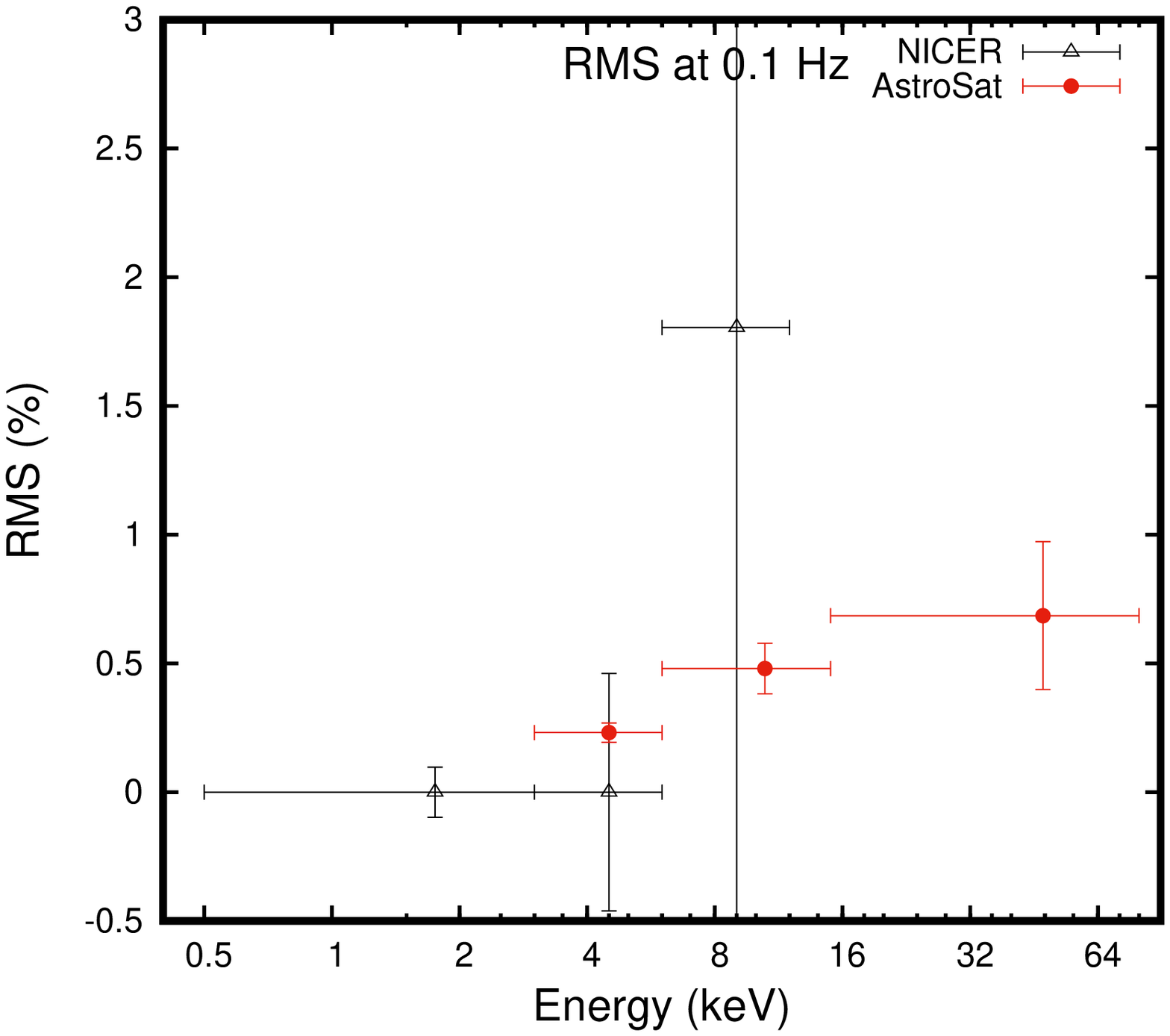}
\includegraphics[width=5.80cm,angle=0]{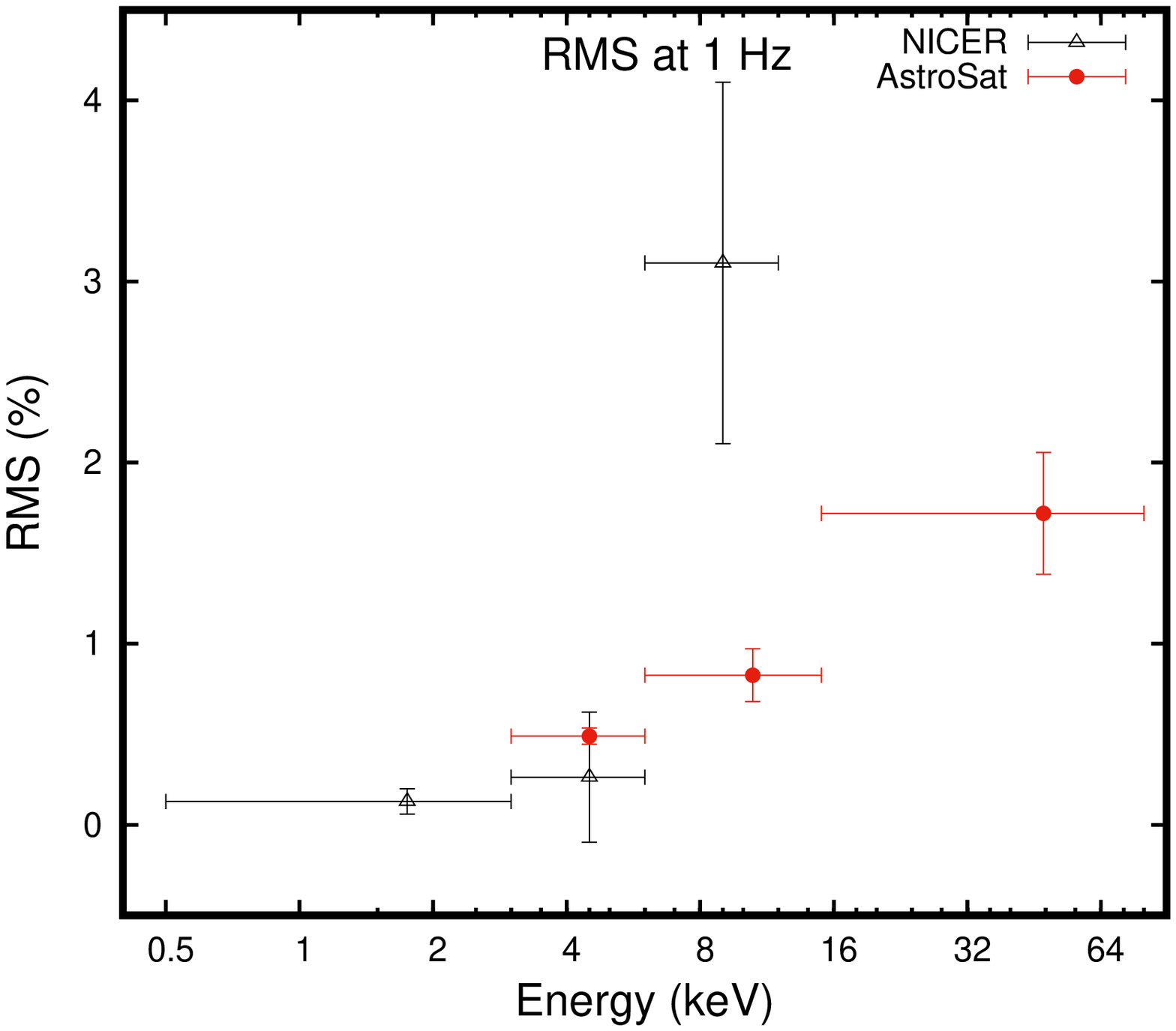}
\includegraphics[width=5.80cm,angle=0]{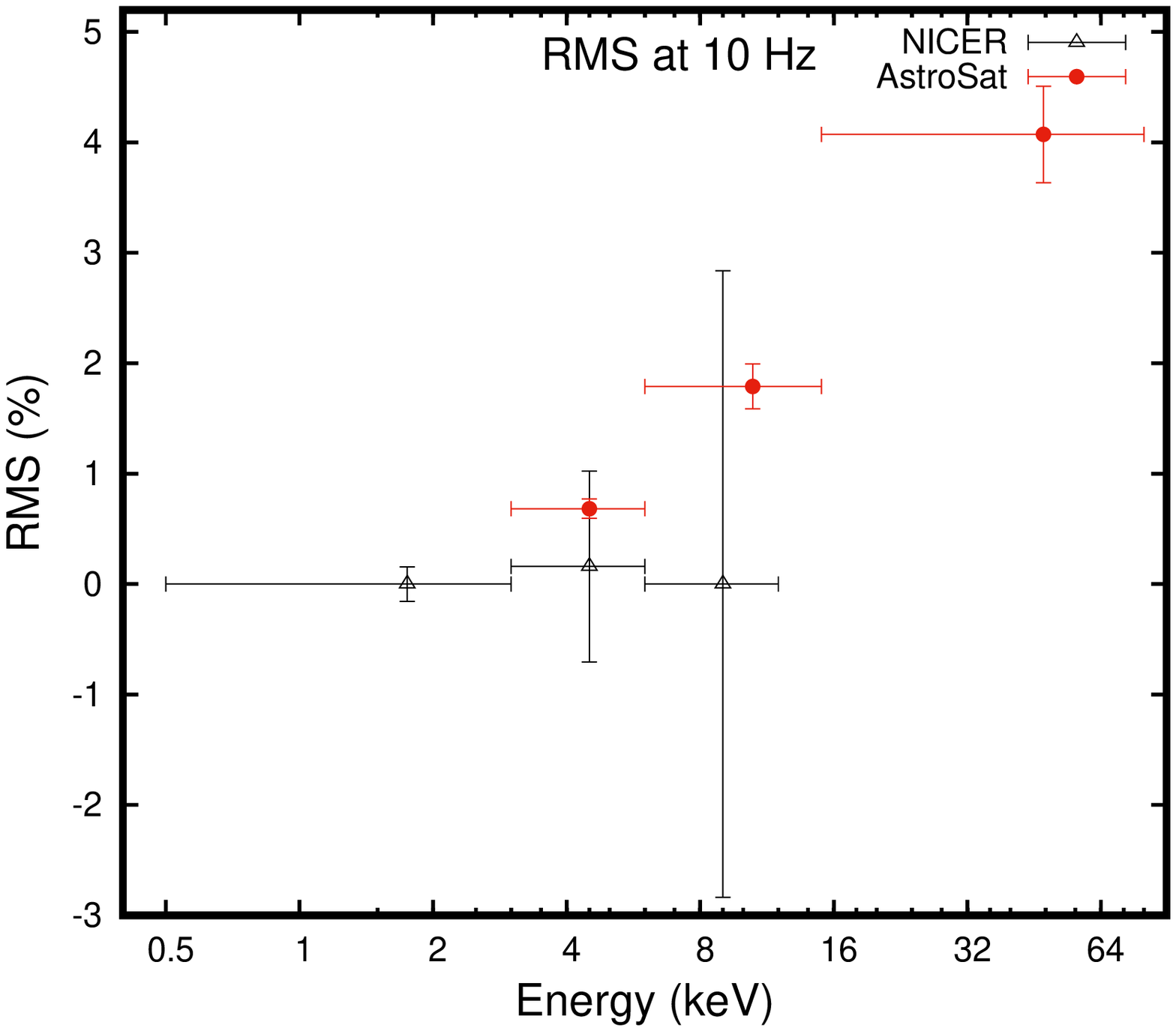}

\caption{Fractional rms as a function of photon energy in the frequency range 0.08--0.12 Hz (around 0.1\,Hz; left), 0.8--1.2 Hz (around 1\,Hz; middle) and 8--12 Hz (around 10\,Hz; right) from the simultaneous {\it NICER} (N1; black open triangle) and {\it AstroSat} (AS1; red filled circle) observations in the soft state.} 
\label{rms_as1_as2_as3}
\end{figure*}

\begin{figure*}
\centering

 \includegraphics[width=8.6cm,angle=0]{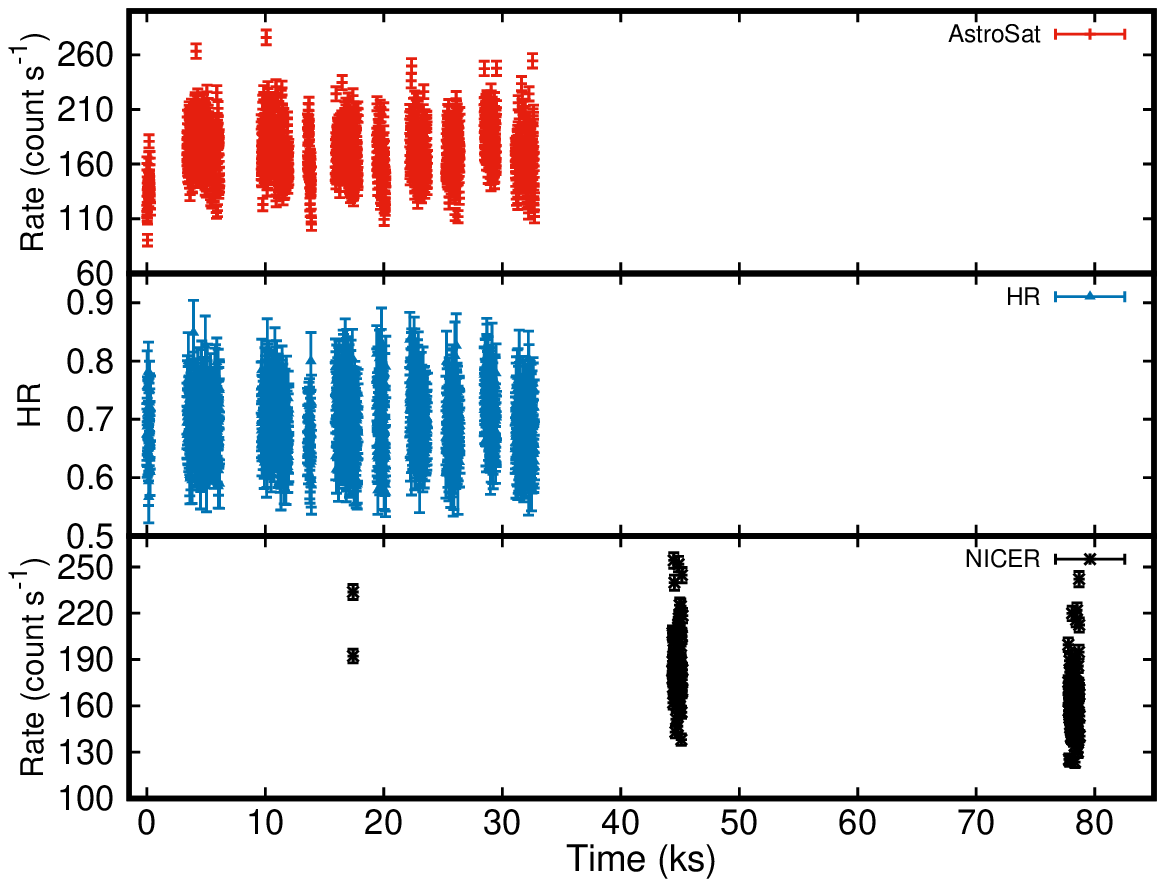}
 \includegraphics[width=8.6cm,angle=0]{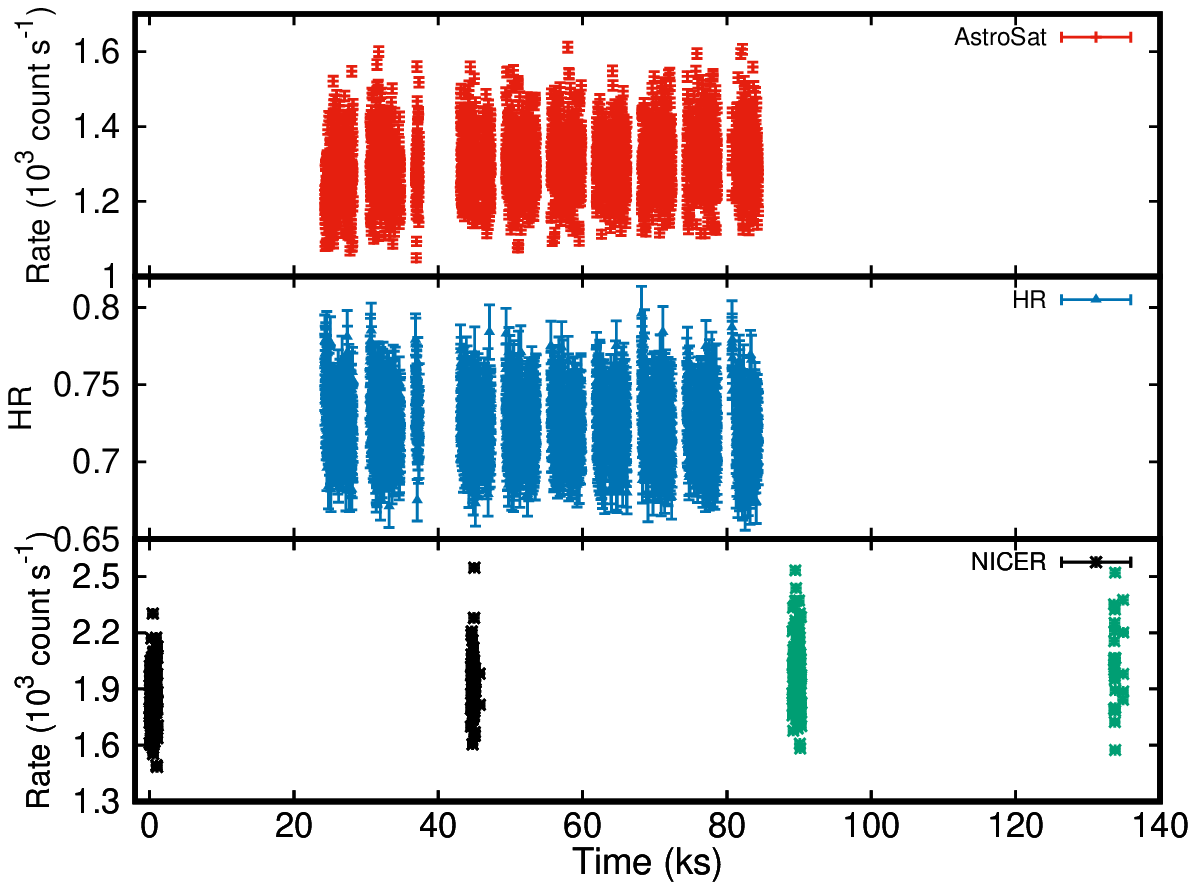}

\caption{The light curves and hardness ratio from simultaneous {\it AstroSat} and {\it NICER} observations in the hard state: AS4 \& N4 (left) and AS5 \& N5 and N6 (right). The labelling is same as that of Figure \ref{lc_soft}.} 
\label{lc_hard}
\end{figure*}

\begin{figure}

 \includegraphics[width=8.75cm,angle=0]{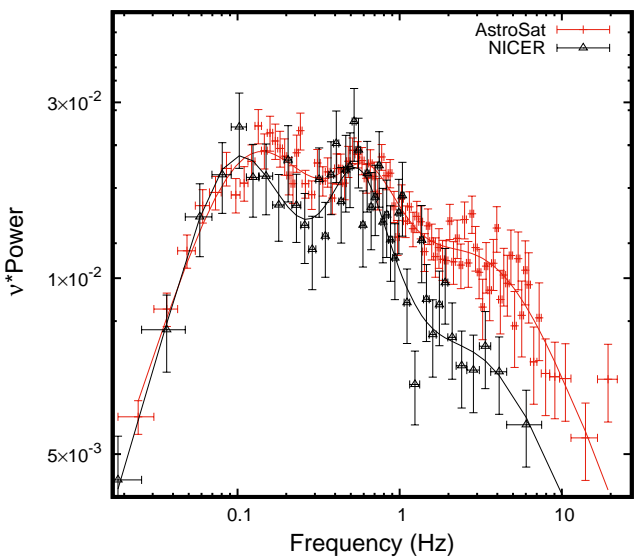}

\caption{PDS from AS4 (red plus points) and N4 (black open triangles) observations in the faint hard state. Both PDS are fitted with three Lorentzians.}
\label{pds_as4}
\end{figure}

\begin{figure*}

\includegraphics[width=8.75cm,angle=0]{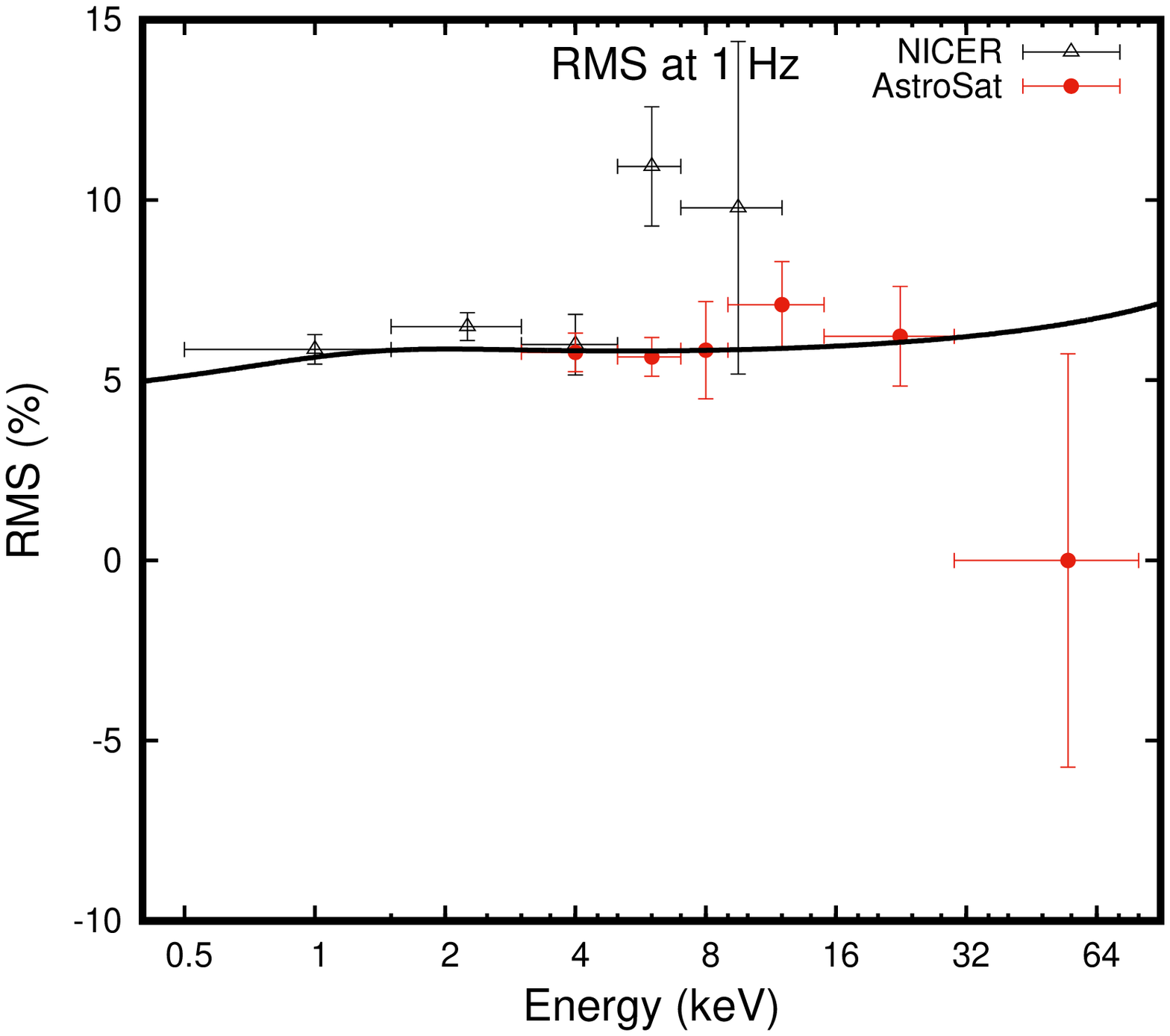}
\includegraphics[width=8.75cm,angle=0]{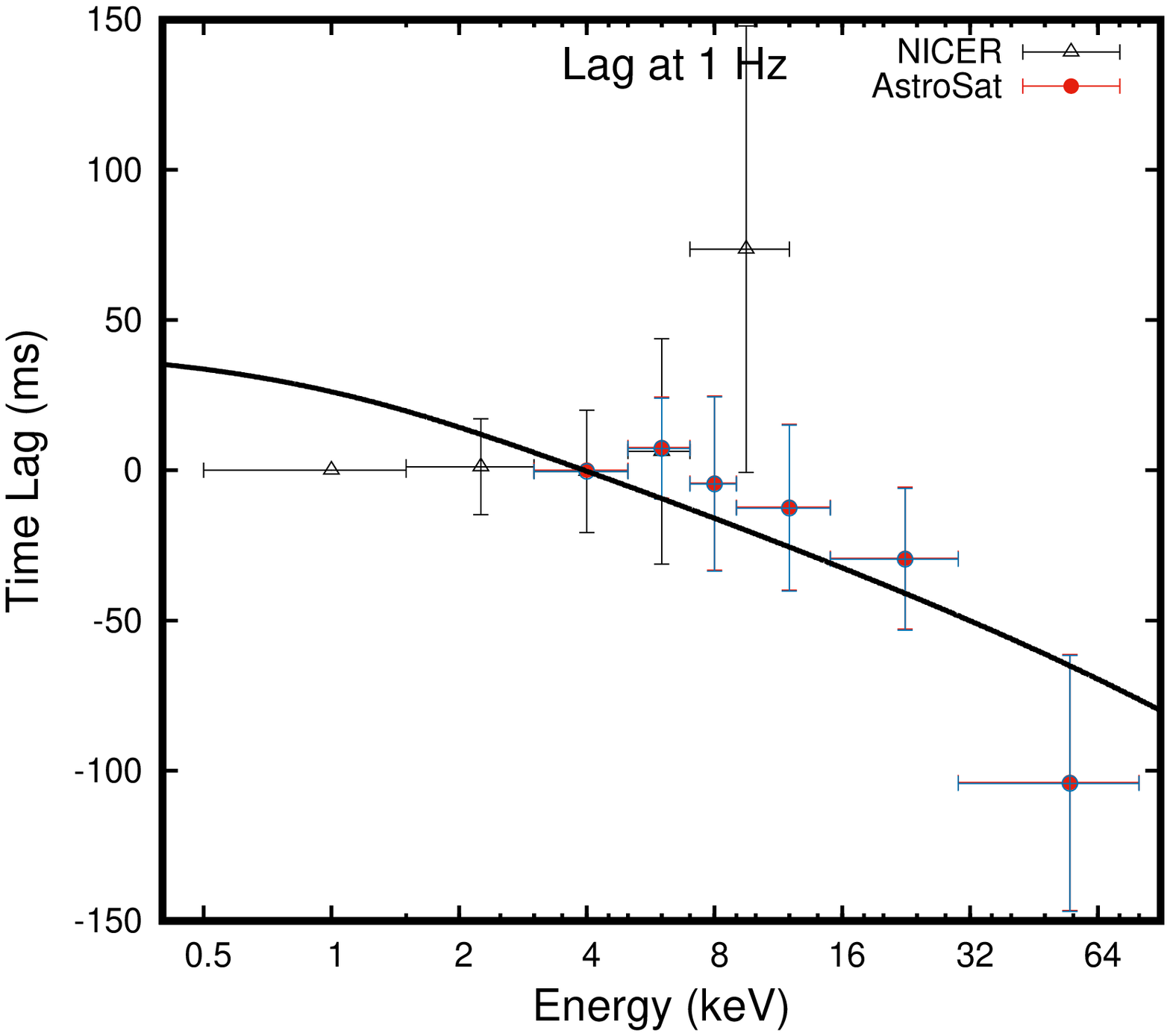}

\caption{Fractional rms (left) and time lag (right) as a function of photon energy in the frequency range 0.8--1.2 Hz (around 1\,Hz) from the faint hard state observation. The red filled circle and black open triangle represent AS4 and N4 observations, respectively. We shifted the time lag detected with {\it AstroSat} to the reference band of {\it NICER}, which is shown in blue open circles. The black solid curve represents the model fit derived from the stochastic propagation model.} 
\label{rms_lag_as4}
\end{figure*}

\begin{figure*}
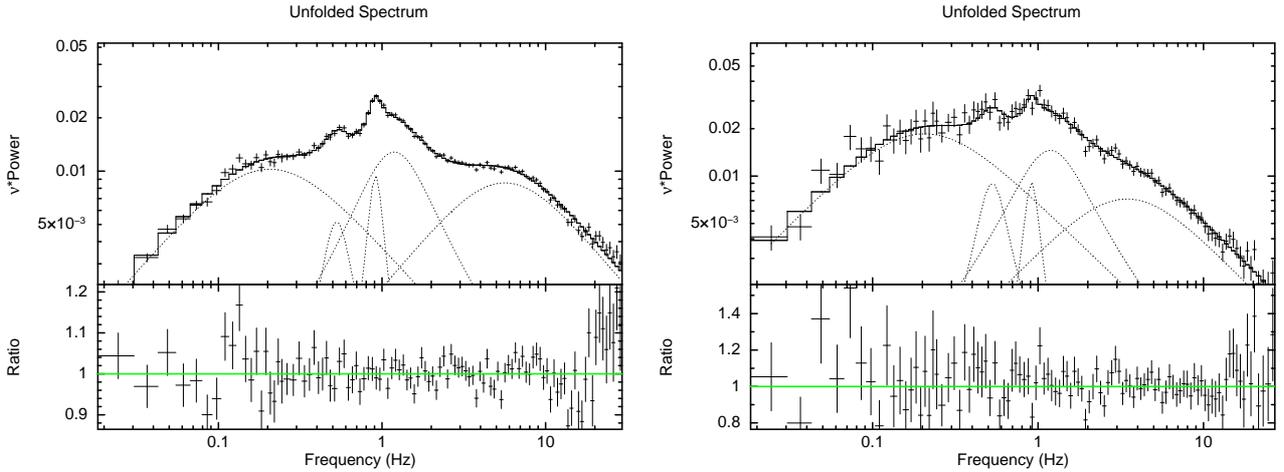


 \includegraphics[width=6.25cm,angle=-90]{f9a.ps}
 \includegraphics[width=6.25cm,angle=-90]{f9b.ps}

\caption{PDS from AS5 (left), N5 and N6 (right) observations in the bright hard state, which are fitted with five Lorentzians. For the {\it NICER} PDS, we fixed the centroid frequency and width of all the Lorentzian functions to the best-fit values obtained from the {\it AstroSat} PDS.}
\label{pds_as5}
\end{figure*}

\begin{figure*}

\includegraphics[width=8.75cm,angle=0]{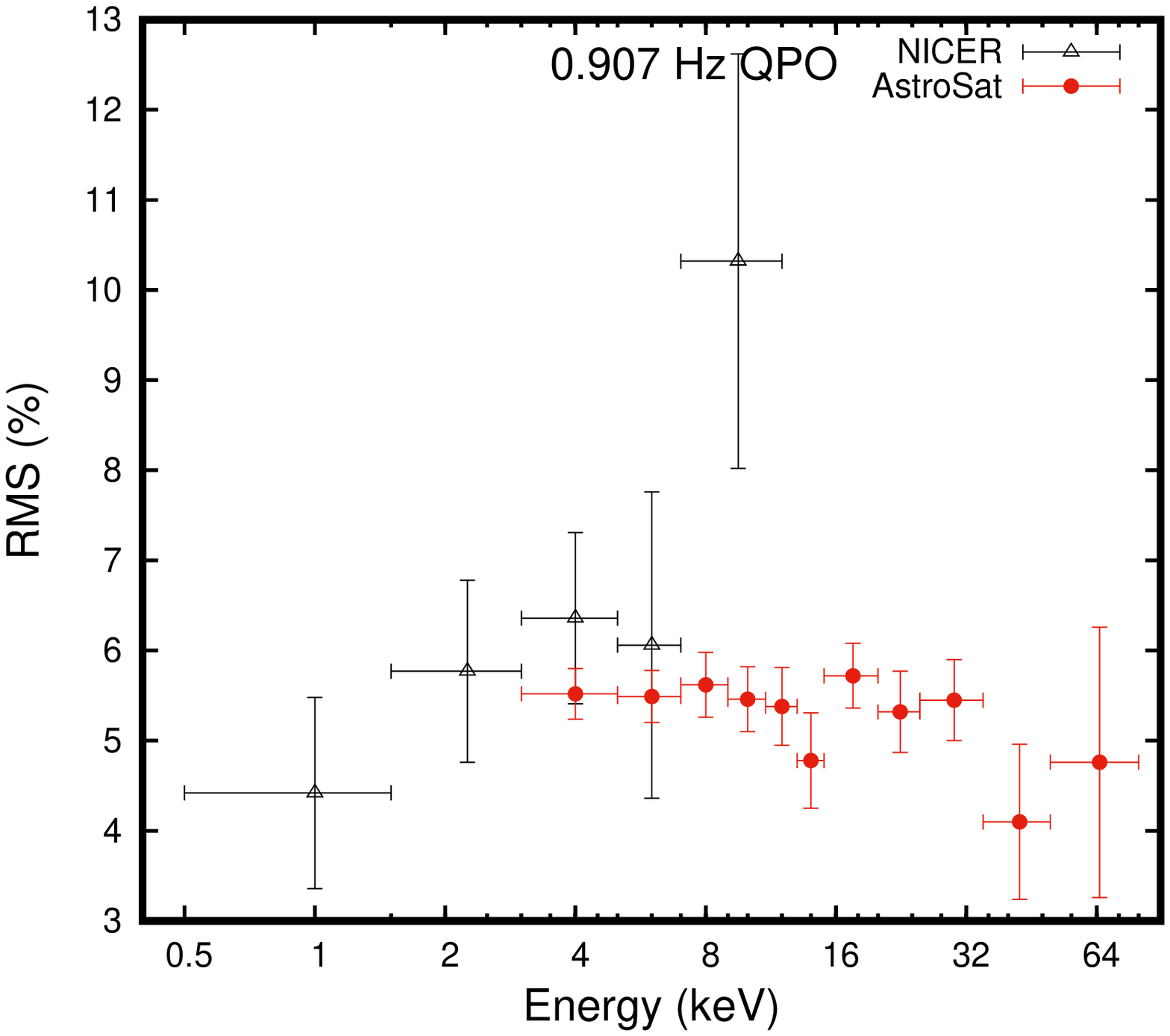}
\includegraphics[width=8.75cm,angle=0]{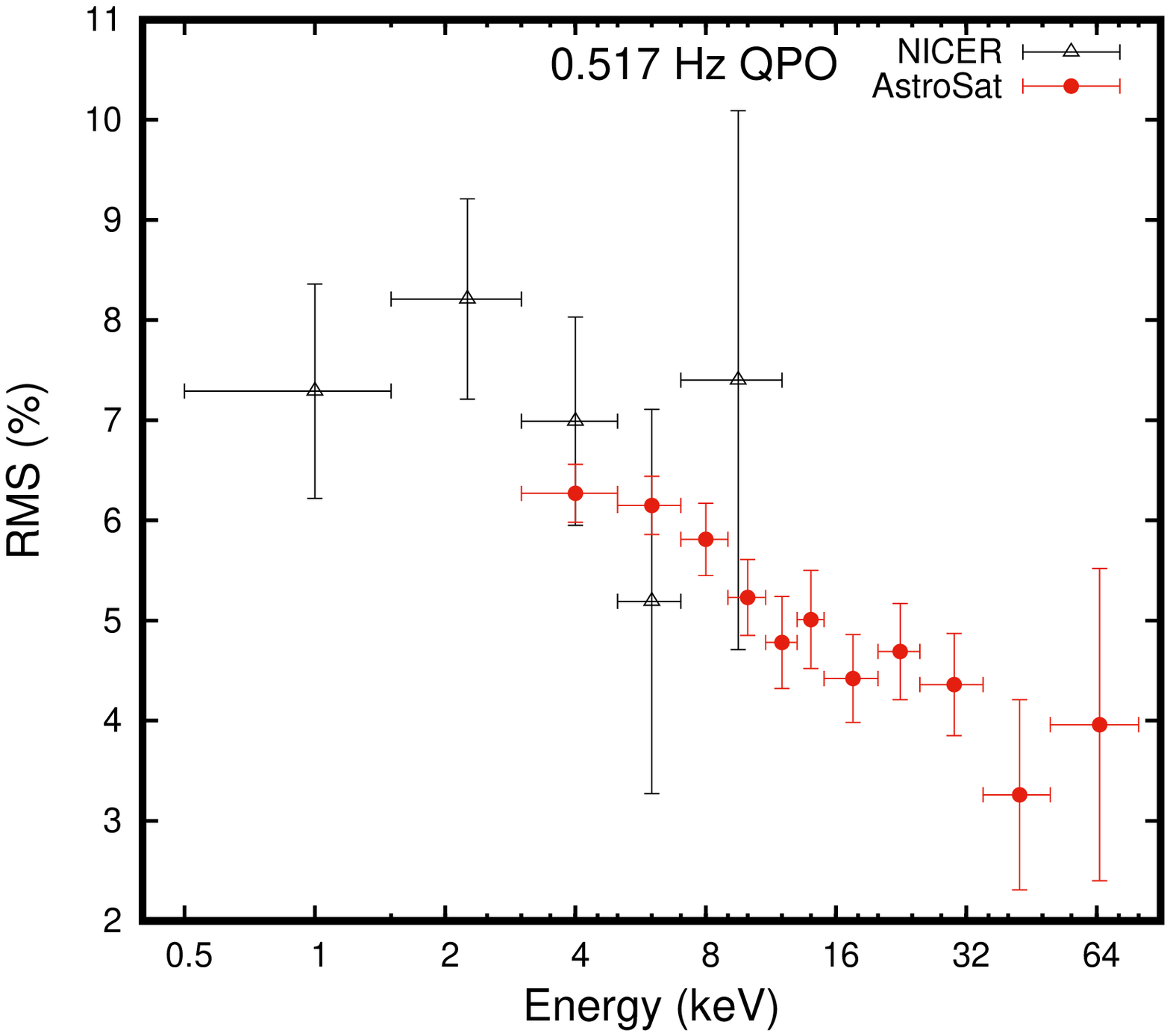}

\caption{Fractional rms as a function of photon energy at 0.907 Hz QPO (left) and 0.517 Hz sub-harmonic (right) from the AS5 (red filled circle) and {\it NICER} (N5 and N6; black open triangle) observations.}
\label{rms_lag_as5_n5_n6}
\end{figure*}

\begin{figure}

\includegraphics[width=8.75cm,angle=0]{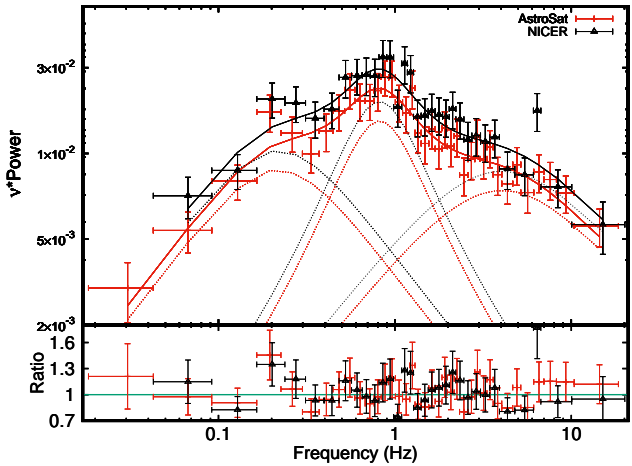}

\caption{The PDS from the strict overlapping time between {\it AstroSat} (AS5) and {\it NICER} (N5) observations in the common energy range of 3--6 keV fitted with three Lorentzians. The red plus points and black open triangles represent {\it AstroSat} and {\it NICER} data, respectively.}
\label{pds_common}
\end{figure}

\begin{figure*}

\includegraphics[width=8.75cm,angle=0]{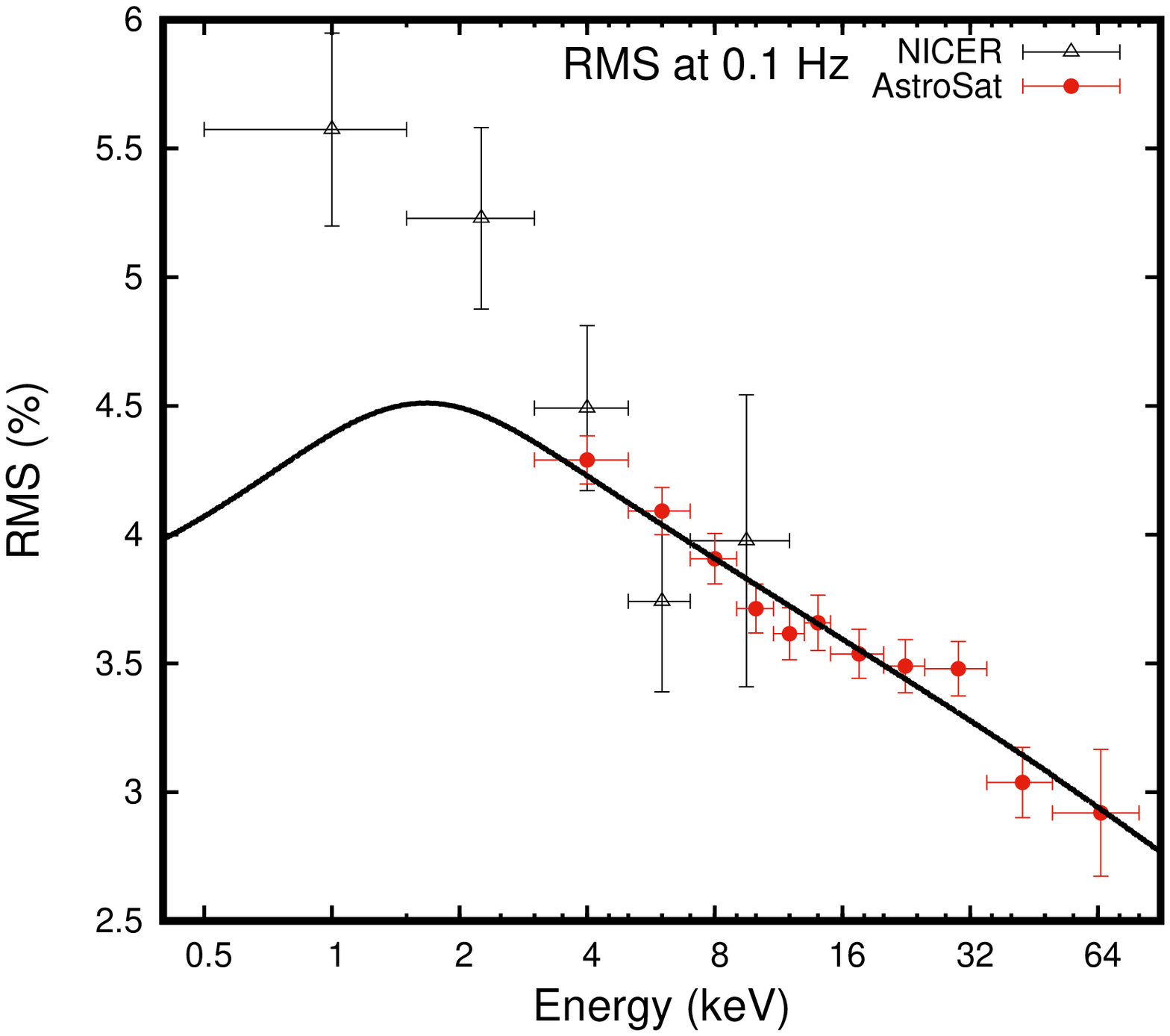}
\includegraphics[width=8.75cm,angle=0]{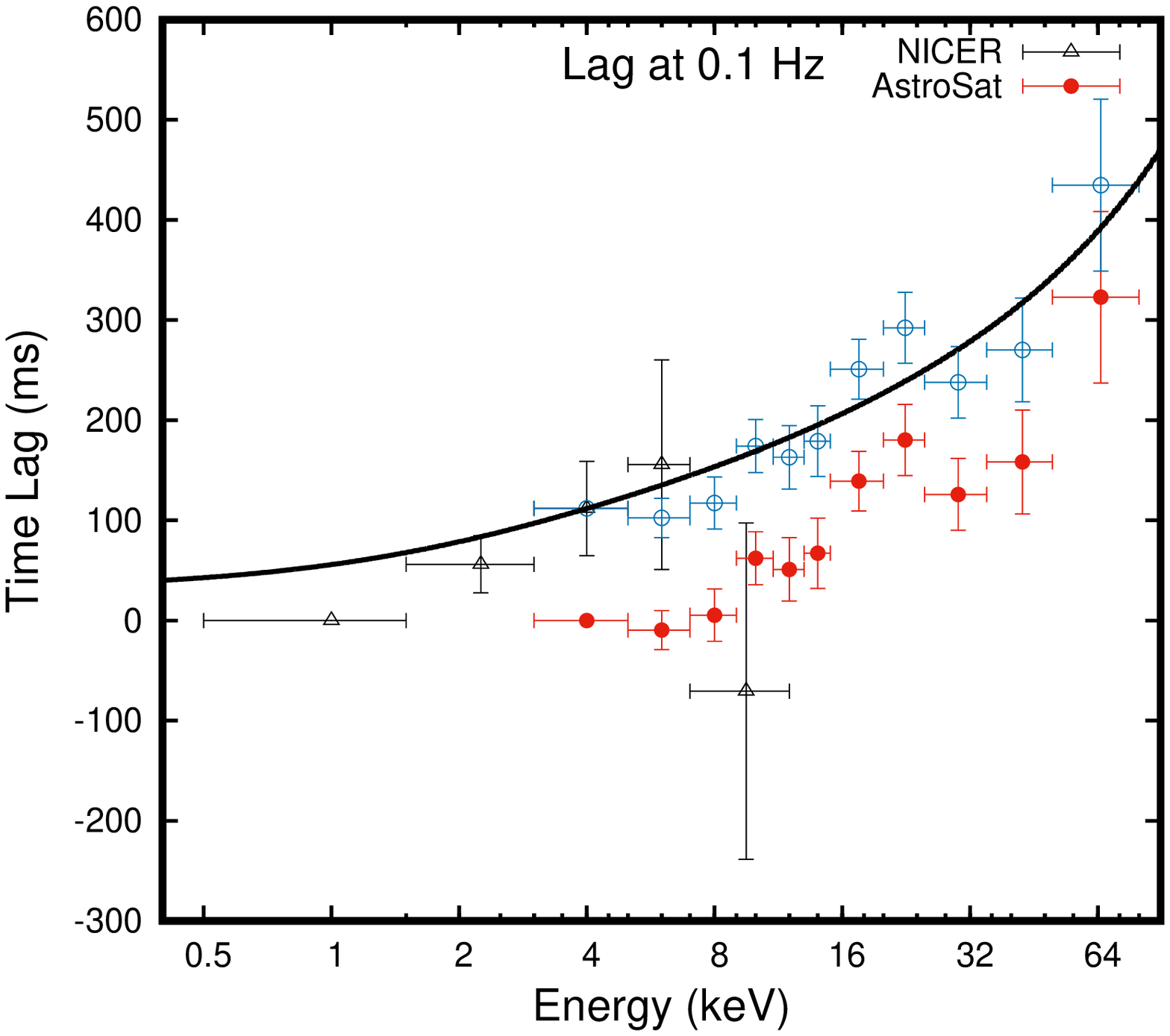}

\includegraphics[width=8.75cm,angle=0]{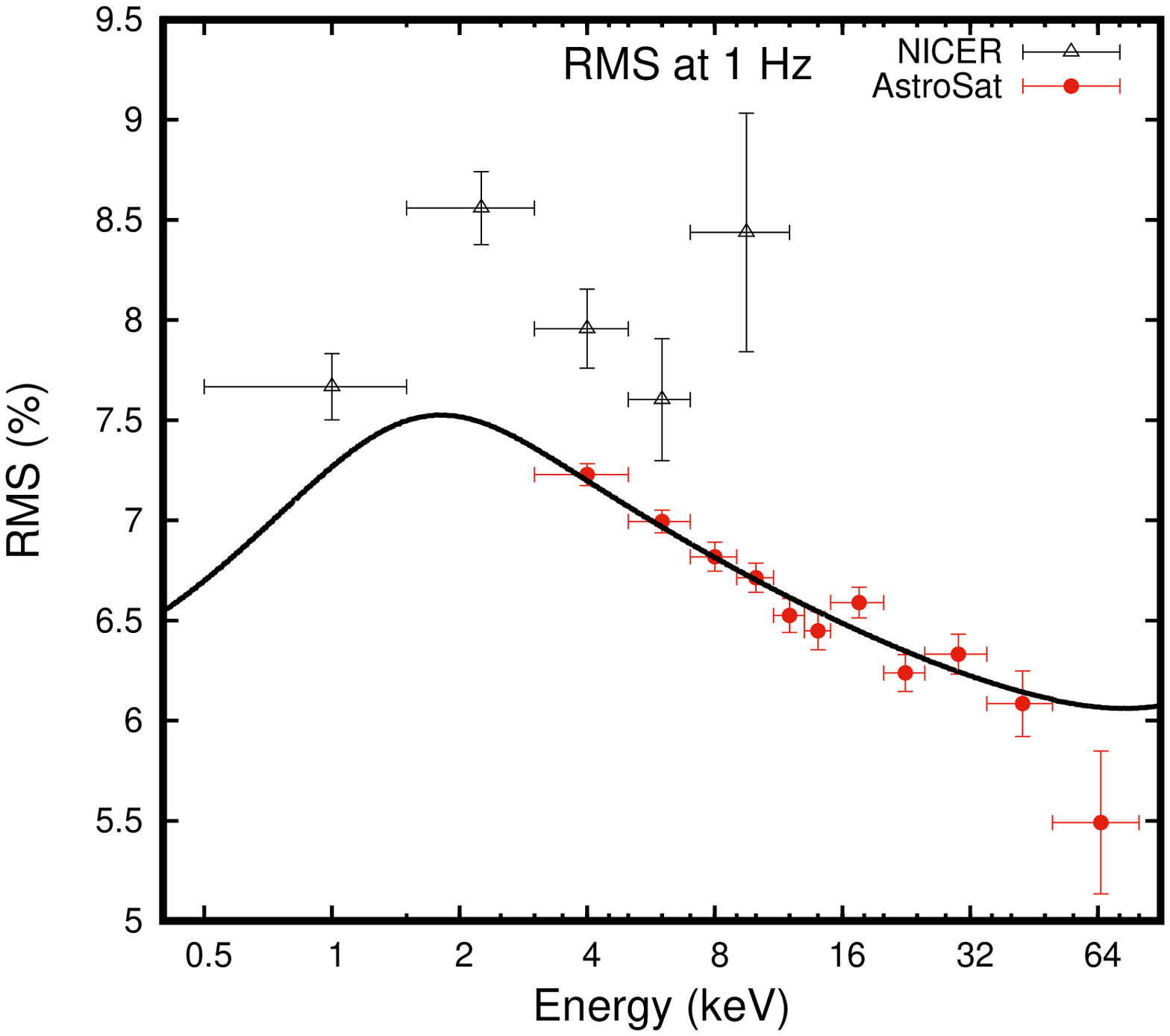}
\includegraphics[width=8.75cm,angle=0]{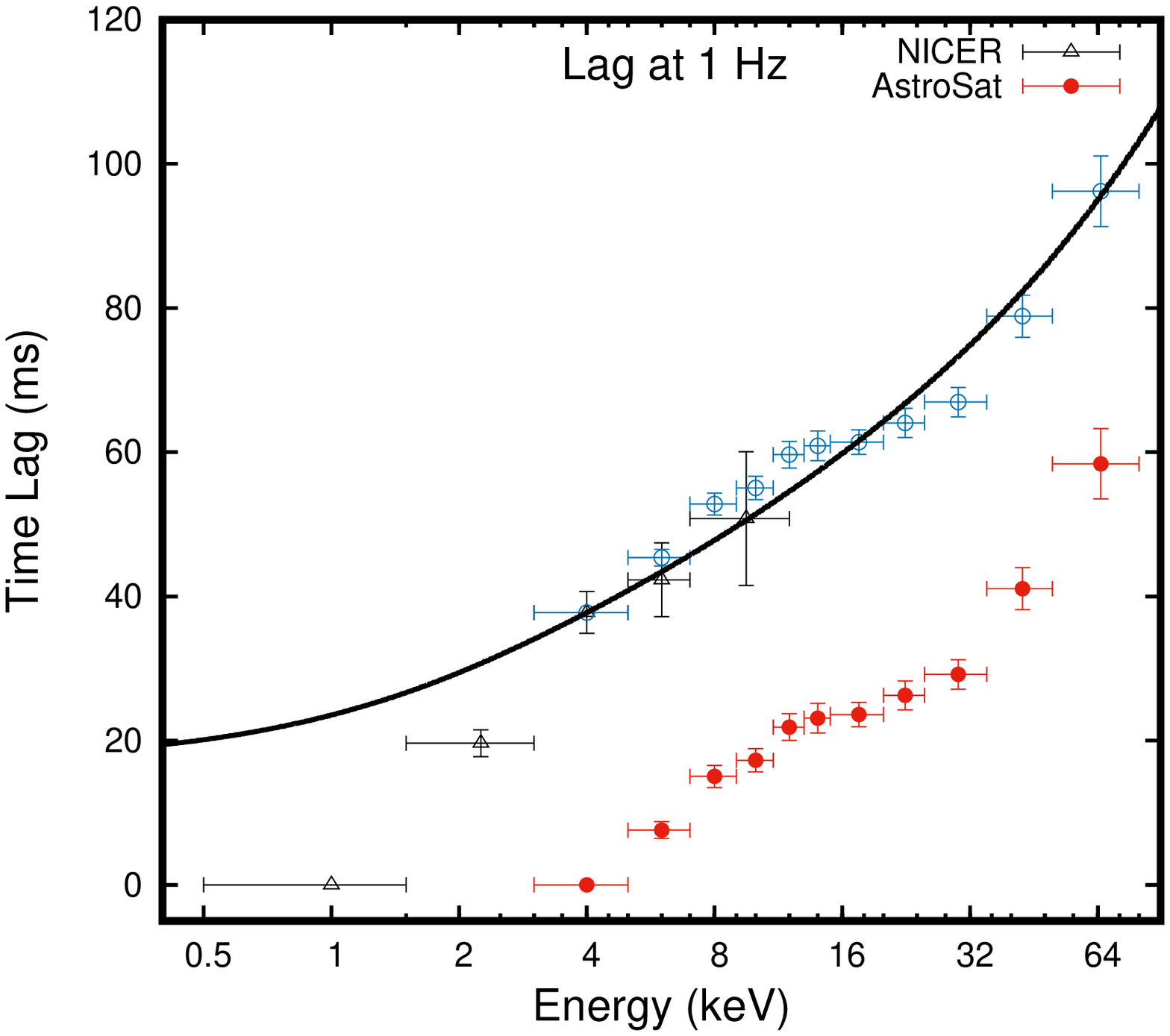}

\includegraphics[width=8.75cm,angle=0]{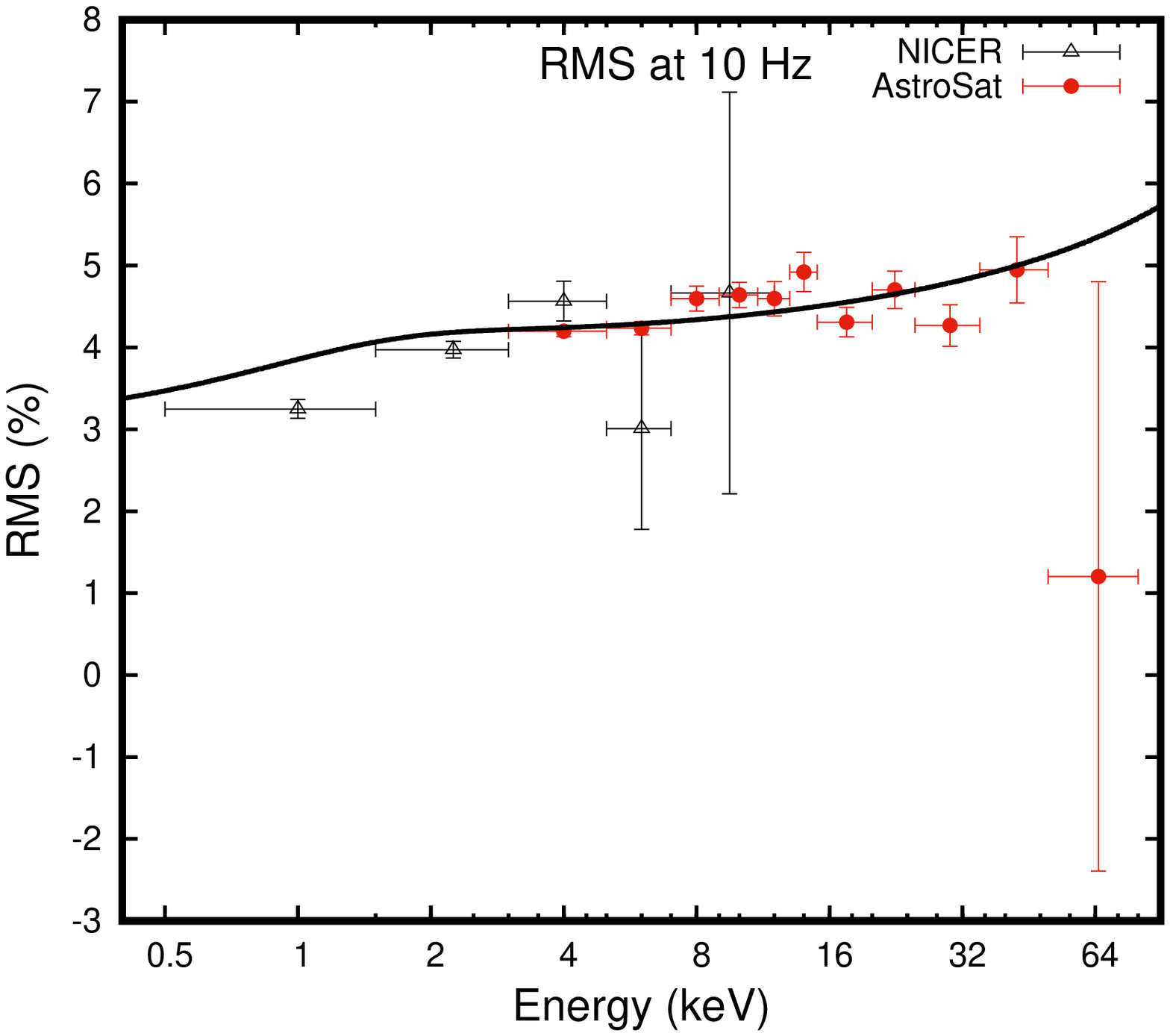}
\includegraphics[width=8.75cm,angle=0]{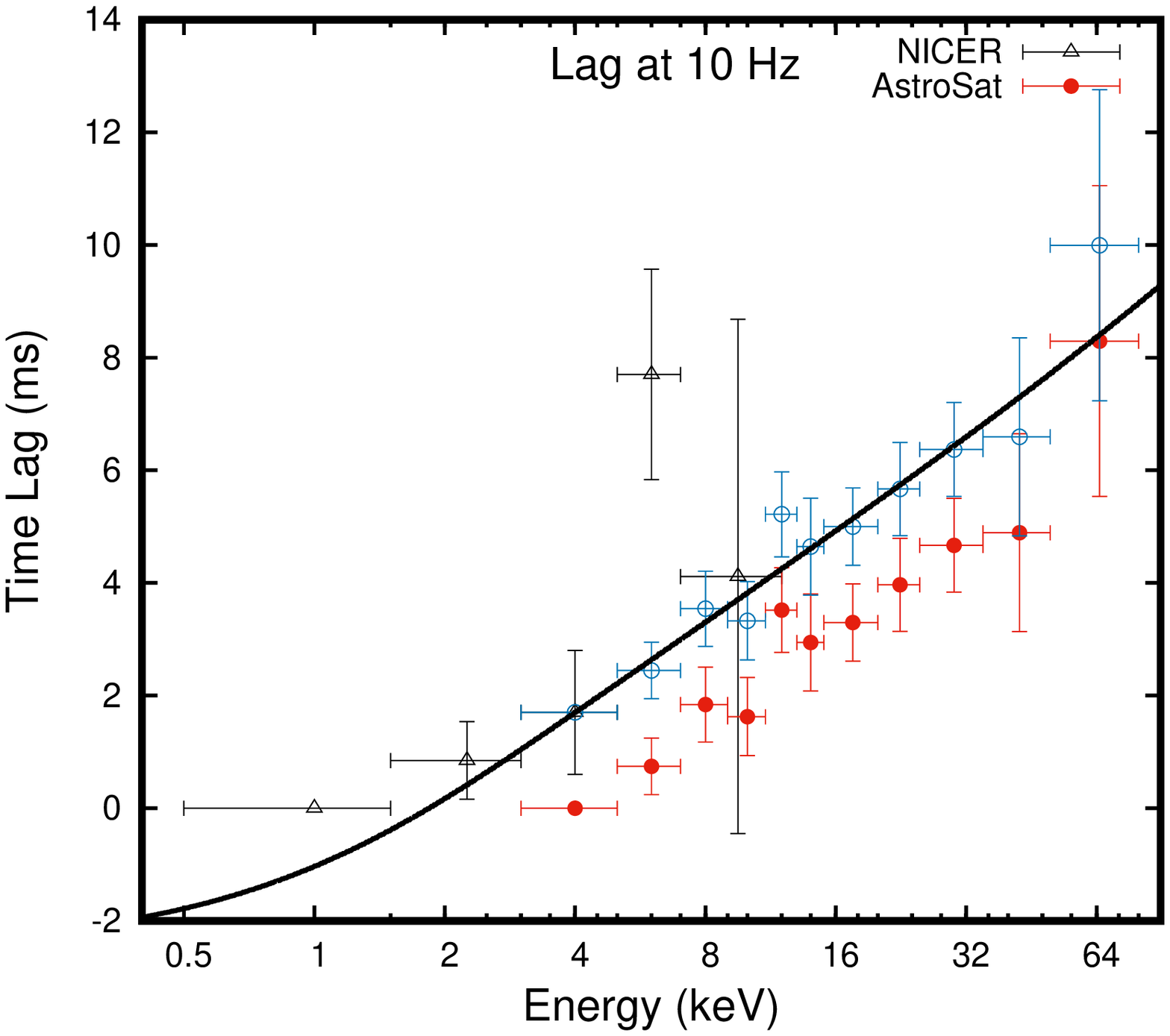}

\caption{Fractional rms (left) and time lag (right) as a function of photon energy at 0.08--0.12 Hz (0.1\,Hz; top panel), 0.8--1.2 Hz (1\,Hz; middle panel) and 8--12 Hz (10\,Hz; bottom panel) frequencies from the simultaneous AS5 (red filled circle) and {\it NICER} (N5 and N6; black open triangle) observations. The LAXPC time lags have been shifted such that the 3--5 keV time lag is same as for {\it NICER} and is shown in blue open circles. The black solid curve represents the model fit derived from the stochastic propagation model.}
\label{rms_lag_as5_n5_n6_3freq}
\end{figure*}

\section{Broadband X-ray Spectral Analysis}
\label{sec:spectral}

We first performed the broadband X-ray spectral analysis of \maxij{} using the SXT and LAXPC instruments onboard {\it AstroSat}. The broadband spectrum in the 0.8--40 keV energy band were modelled with {\sc xspec} version 12.10.1f \citep{Arn96}. Here our motivation is to understand the spectral state of source and hence we used a simple model for the broadband spectral modelling, which consists of a multi-colour disk blackbody \citep[MCD; {\tt diskbb} in {\sc xspec};][]{Mit84} and a Comptonization component \citep[{\tt simpl};][]{Ste09} along with a gaussian line profile ({\tt gaussian}). Detailed spectral modelling taken into account relativistic effects will be presented elsewhere. We used Tuebingen-Boulder Inter-Stellar Medium absorption model \citep[{\tt tbabs};][]{Wil00}. A multiplicative constant was used to address the cross-calibration uncertainties between the SXT and LAXPC instruments. A 3\% model systematic uncertainty is used for spectral modelling. The parameter errors are at a 90\% confidence level.

The unfolded spectra and residuals from the five observations are depicted in Figure \ref{5spec} and the best-fit model parameters are listed in Table \ref{spec_params}. The absorption column density seems to be a constant at $\sim 5 \times 10^{21}~\rm cm^{-2}$ for these observations, except in AS4 observation, where it dropped to $\sim 4 \times 10^{21}~\rm cm^{-2}$. The photon index, scattering fraction and inner disk temperature significantly changed during these observations. The photon index decreases from $\sim 2.1$ to $\sim 1.6$, while the scattering fraction increases from $\sim 7\%$ to $\sim 50\%$. The inner disk temperature shows a decreasing trend from $\sim 0.8$ to $\sim 0.3$ keV. The unabsorbed flux was computed using the convolution model {\tt cflux} in the 0.8--40 keV energy band. The total flux increases from $\sim 4$ to $\sim 4.7 \times 10^{-8}\rm~erg~cm^{-2}~s^{-1}$ in the first three observations. In the AS4 observation, the flux drops by a factor $\sim 15$ compared to AS3 observation and then increases to $\sim 2 \times 10^{-8}\rm~erg~cm^{-2}~s^{-1}$ in the last {\it AstroSat} (AS5) observation.

In the first three observations (AS1, AS2 and AS3), the value of power law index is $> 2$, the inner disk temperature $\sim 0.8$ keV, the scattered fraction of seed photon from the accretion disk is $\lesssim 10 \%$ and estimated disk fraction is $> 75\%$, which identifies the source to be in the soft spectral state of BHXRBs. The X-ray spectrum of the source significantly changed in the last two observations (AS4 and AS5), with photon index $\sim 1.6$, the inner disk temperature dropped to $\sim 0.3$ keV, the scattered fraction of $\sim 20-50\%$ and disk fraction dropped below 15\%. Based on the spectral parameters, we have identified the source to be in the hard X-ray spectral state of BHXRBs \citep{Rem06, Bel10, Bel11} in these observations. In the next section, we discuss the broadband time variability. We first consider the three soft state observations and then the two hard state ones.

\section{Broadband Timing Analysis}
\label{sec:timing}

\subsection{Soft State Observations}

The {\it AstroSat} LAXPC20 background subtracted 3--80 keV light curves of \maxij{} in the soft state (AS1, AS2 and AS3) are depicted in the top panels of Figure \ref{lc_soft}. In the AS2 observation, the intensity increased from $\sim 3000~\rm count~s^{-1}$ to $\sim 3400\rm~count~s^{-1}$ in a short period and then the source flipped back to the lower intensity level. The hardness ratio (HR) is defined
as the ratio between the 7--16 keV  and the 3--7 keV rate and its variation is shown in the middle panels of Figure \ref{lc_soft}. The 0.5--12 keV light curves from {\it NICER} observations, which are simultaneous to {\it AstroSat} are plotted in the bottom panels.

The power density spectrum (PDS) in the 0.01--30 Hz frequency range from LAXPC and {\it NICER} data for the energy range 3--15 keV and 0.5--12 keV, respectively, are shown in Figure \ref{pds_first3}. The LAXPC PDS shows complex broad band features and require to be fitted by multiple Lorentzians. Since the source exhibited different intensity levels in the LAXPC light curve of AS2 observation,  we extracted the PDS from low and high-intensity levels which are shown in the middle panel of Figure \ref{pds_first3}. The PDS from the low and high intensity levels are different and the PDS becomes weaker in the high-intensity level compared to the lower one. In addition, we can see a broad feature (the centroid frequency is $\sim 4.6$\,Hz and the width is $\sim 4.3$\,Hz) appeared in the PDS extracted from the high intensity level. In the PDS of AS3 observation, we can see a QPO at $6.85^{+0.24}_{-0.26}$\,Hz with a width $3.80^{+0.85}_{-0.74}$\,Hz. The detected QPO at $\sim 6.9$ Hz is broad (the $Q$-factor is $\sim 1.8$) and the rms is $< 2\%$ in the {\it AstroSat} energy bands.  

In contrast, the PDS obtained from {\it NICER} data shows significantly lower values and hence can be modelled using a simple power-law as shown in Figure \ref{pds_first3}. The difference in variability between the {\it NICER} and LAXPC suggests a strong energy dependence of the fractional rms. To verify if that is the case, we extracted the fractional rms, by following the methods discussed in \citet{Now99}, as a function of energy, for three frequency ranges (0.08--0.12, 0.8--1.2 and 8--12 Hz) for the first observation (AS1 and N1) and plotted them in Figure \ref{rms_as1_as2_as3}. We note that for the common energy range of 3--5 keV the {\it NICER} and LAXPC fractional rms are consistent with each other. The variability is more pronounced at higher energies reflecting a larger fractional rms in the LAXPC bands. We have also computed the time lag at three frequency ranges using several energy bands 
from {\it NICER} and {\it AstroSat}. To compute the 
time lag, we used 0.5--3 keV ({\it NICER}) and 3--6 keV 
({\it AstroSat}) as the reference energy bands in the 
AS1, AS2, AS3 and corresponding NICER observations. We 
do not observe any trend in the time lag spectra, 
hence those lag spectra are not shown in the paper.

In \S \ref{sec:spectral}, spectral analysis of the soft state data revealed that the disc emission dominates for energies $< 4$ keV as is seen in the top three panels of Figure \ref{5spec}. Thus, the prominent variability observed in LAXPC data and a significantly reduced variability in the {\it NICER} one, can be understood in the framework where the disk emission is non-variable while the Comptonized component rapidly varies. While this has been inferred from earlier temporal analysis of black hole systems in the soft state, here we confirm the results using broadband spectral and timing analysis.

\subsection{Hard State Observations}

For the two hard state observations (AS4, AS5 and corresponding {\it NICER} observations) the intensity varied significantly between the observations, although the hardness ratio remained similar for both (see Figure \ref{lc_hard}). There were no significant flux variations seen in the light curves of each of the observations. The difference in the intensity prompted us to name the data set AS4 and N4 as belonging to the faint hard state and the AS5 and corresponding {\it NICER} observations as the bright hard state. We study the rapid variability of these two states separately in the next two sub-sections.

\subsubsection{Faint Hard State}

The power density spectra for the faint hard state observations show broad features in the 0.01--30 Hz frequency range with no QPOs. In contrast to the soft state observations described in the previous section, the PDS are  similar in the low energy ({\it NICER}) and high energy (LAXPC) data. Figure \ref{pds_as4} shows the PDS generated from LAXPC data in the 3--15 keV along with that generated from {\it NICER} data in the 0.5--12 keV range. Three broad Lorentzians have been used to model both the data sets.

To explore the energy dependence further, we extracted the fractional 
rms at three frequency ranges 0.08--0.12, 0.8--1.2 and 8--12 Hz, as was
done for the soft state observations. The rms seems to be a constant 
around $\sim 7\%$ and the time lags are consistent with zero at $\sim 0.1$ 
and $\sim 10$ Hz. Thus, we show the rms and lag spectrum for the frequency 
range  0.8--1.2 Hz, from both {\it NICER} and LAXPC data in Figure \ref{rms_lag_as4}.
The variability seems to be nearly a constant at $\sim 7$\% over all energies. There is a slight discrepancy at the common energy range of 5--7 keV for {\it NICER} and LAXPC data, but the deviation is within two sigma and moreover the data are not strictly simultaneous (see Figure \ref{lc_hard}). The time lag versus energy for the same frequency range is shown in the right panel of Figure \ref{rms_lag_as4}. Here the reference energy band for the NICER data is 0.5--1.5 keV while for LAXPC it is  3--5 keV. While there is no significant time lag for the {\it NICER} data, there is a soft lag at high energies for LAXPC data of the order of 100 ms. Such a soft lag is rather unusual for the hard state of black hole binaries.

\subsubsection{Bright Hard State}

After the main outburst, the source exhibited a re-flare which peaked around MJD 58648 (see Figure \ref{maxi-lc}). The source's intensity increased to $\sim 1300\rm~count~s^{-1}$ in the bright hard state observation (AS5) compared to the earlier hard state observation where the count rate was $\sim 200\rm~count~s^{-1}$. The higher count rate allows for a more detailed timing analysis.

The PDS obtained from LAXPC in the 3--15 keV energy band is shown in the left panel of Figure \ref{pds_as5}. The PDS can be described by five Lorentzian functions which represent three broadband components along with a QPO at $0.907\pm0.01$ Hz ($Q$-factor $\sim 4.7$) and a possible sub-harmonic at $0.517\pm 0.02$ Hz ($Q$-factor $\sim 2.4$) with widths $0.194\pm0.06$ Hz and $0.223\pm0.07$ Hz, respectively. Here, the Lorentzian component with the higher $Q$-factor is considered to be the primary QPO, while the other one is taken to be the sub-harmonic. Indeed, the temporal behaviour of the systems at $\sim 1$ Hz is complex, with a broader component peaking at $\sim 1.5$ Hz close to the QPO feature. The fitting resulted in a formal $\chi^2$/dof$= 160.6/93$. 

The right panel of Figure \ref{pds_as5} shows the PDS generated from the {\it NICER} observations in the energy range 0.5--12 keV. The overall shape of the PDS is similar to that obtained from LAXPC but with changes in relative strengths of the components. This is demonstrated by fitting the same five Lorentzian as used for LAXPC data for the {\it NICER} one. Here, the centroid frequency and width of the Lorentzian components are fixed to the values obtained for the LAXPC fitting, but allowing for the normalization to vary. This leads to an acceptable $\chi^2$/d.o.f $= 93.3/98$. While the QPO components seems to have nearly the same strength for both observations, the sub-harmonic component normalization is lower for LAXPC data. To quantify the energy dependence of the QPO features, the PDS at different energies were fitted with the same five component model and the normalization of the Lorentzians was used to determine the fractional rms. The results are illustrated in Figure \ref{rms_lag_as5_n5_n6}, where the left panel shows that strength of the primary QPO is nearly energy independent, while for the sub-harmonic it decreases with energy.

The slight difference in the temporal behaviour seen between the {\it NICER} and LAXPC analysis could be either due to the different energy band used for the two instruments or it could be that the source behaviour changed for the two observations, since they are not strictly simultaneous. To check for these possibilities, PDS was generated for the small $\sim 500$ seconds strict simultaneous data available for both instruments (see right panel of Figure \ref{lc_hard}) in the common energy range of 3--6 keV. Figure \ref{pds_common} shows the two strict simultaneous PDS, which shows that the overall shape of the PDS from the two instruments are remarkably similar, with the {\it NICER} data showing a slight higher variability. The slightly lower PDS values for the LAXPC data may be due to dead time effects. While dead time corrections have been incorporated in the Poisson level of the LAXPC data \citep{Yad16a}, the real variability strength may be slightly smaller due to dead time which has not been taken into account here. Both the PDS can be represented by three Lorentzian components. Unfortunately, the QPOs and the complex features seen at 1 Hz, are not distinguishable here for this short time observation. Thus, while Figure \ref{pds_common} illustrates how well the analysis of two instruments agree with each other, it is not clear whether the slight variation seen in the PDS in Figure \ref{pds_as5} for the LAXPC and {\it NICER} data is due to energy dependence or source variation. Nevertheless, we continue with further energy-dependent temporal analysis of the data keeping in mind that the data is not strictly simultaneous.

\begin{figure}

\includegraphics[width=8.75cm,angle=0]{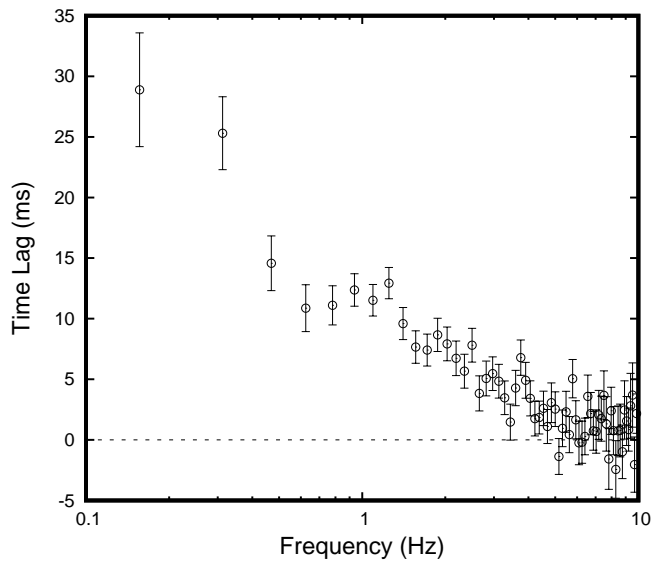}

\caption{Frequency-dependent time lags between 3--4 keV and 4--12 keV band light curves from AS5 observation.}
\label{lag-fre}
\end{figure}

Since the statistics are not good enough to ascertain the detailed energy-dependent properties such as time lag for the QPO and other individual features we study the energy-dependent temporal properties of the source in three broad frequency bands. Figure \ref{rms_lag_as5_n5_n6_3freq} shows the fractional rms and time lag as a function of energy for the frequency ranges 0.08--0.12 Hz (around 0.1\,Hz), 0.8--1.2 Hz (around 1\,Hz) and 8--12 Hz (around 10\,Hz). The reference energy bands for the time lag computation is 0.5--1.5 keV and 3--5 keV for the {\it NICER} and LAXPC bands, respectively. Also shown in the right panel (with blue open circles) the LAXPC time lag shifted such that the 3--5 keV time lag is the same as that observed by {\it NICER}. Hence these data points can be considered to be having the same reference energy band as that of {\it NICER} i.e. 0.5--1.5 keV. To further understand the frequency-dependent time lag, we extracted the frequency-dependent time lags between 3--4 keV and 4--12 keV energy band for the bright hard state observation and shown in Figure \ref{lag-fre}. A hard time lag has been observed at all frequencies in this observation, which is consistent with time lag spectra extracted from three frequency ranges.

Broadly, the rms is seen to decrease with energy for $\sim 0.1$ and $\sim 1$ Hz, while for $\sim 10$ Hz there is a marginal increase with energy. The magnitude of the time lags depend on the frequency range and are found to be hard lags i.e. the lags increase with energy. Note the apparent difference in nature of the rms versus energy when a broad frequency range is taken around 1 Hz (right middle panel of Figure \ref{rms_lag_as5_n5_n6_3freq}) as compared to when it is estimated only for the QPO at 0.91 Hz (left panel of Figure \ref{rms_lag_as5_n5_n6}), although the errors are larger for the QPO. This indicates the complexity of the temporal behaviour of the system at around 1 Hz. 

In the next section we model and interpret the energy-dependent rms and time lag at different frequencies in terms of a single-zone stochastic propagation model applicable for the hard state of black hole binaries.

\section{Modelling the Energy Dependent Timing Properties}
\label{sec:modelling}

One of the promising models to explain the timing properties of BHXRBs is the stochastic propagation one \citep{Lyu97, Kot01, Ing11, Ing12, Ing13}. In this interpretation variability induced in the outer regions of the disk, at different frequencies, propagate inwards to produce the observed variability in the X-ray band. Different versions of this generic scenario can be invoked to predict the fractional rms and time lag as a function of energy. \citet{Maq19} have described in detail a single-zone stochastic propagation model which they used to fit the energy-dependent timing properties of Cygnus X--1. This model has also been successfully used to  fit the energy-dependent timing features of the broadband noise in MAXI J1820+070 \citep{Mud20} and of the QPO in Swift J1658.2$-$4242 \citep{Jit19}.

While details of the single-zone propagation model are presented in \citet{Maq19}, here we briefly mention some of the primary components. The assumed geometry of the system is a truncated standard disk characterised by a inner disk temperature $T_{\rm in}$, with a hot inner flow having a single uniform temperature $T_{\rm e}$, hence the model is termed as a single-zone one. The inner flow Compontonizes photons from the truncated disk to produce the observed hard X-ray emission.  Variations in the inner disk temperature $\delta T_{\rm in}$, changes the input photons inducing a variation in the Comptonized spectrum. Additionally, there is a variation in the heating rate of the hot inner flow inducing a variation in its temperature $\delta T_{\rm e}$, which can occur after a time delay compared to $\delta T_{\rm in}$.

The model requires the parameters of the Comptonized component obtained from the time averaged spectrum, namely the inner disk temperature, the power-law index and the temperature of the hot inner region. The first two are estimated directly by the spectral fitting described in Section \ref{sec:spectral} (see Table \ref{spec_params}). Since the temperature of the hot inner region cannot be constrained by the spectral fitting we assume it to be $100$ keV. The other parameters required are the variation $\delta T_{\rm e}$, the ratio $\delta T_{\rm in}/\delta T_{\rm e}$ and the time delay between them $\tau_D$.

Since the model is applicable only to the thermal Comptonized component, we restrict the analysis only to the hard state data. Furthermore, we only formally fit the LAXPC data in the energy range (3--80 keV) and extrapolate the predicted variability to low energies to compare with the {\it NICER} results. The best-fit parameters are listed in Table \ref{modelfit} and the model is plotted as black solid line in Figures \ref{rms_lag_as4} and \ref{rms_lag_as5_n5_n6_3freq}. For the time lag curves, the model prediction are shifted such that the time lag at 3--5 keV matches with the observed {\it NICER} values and hence can be interpreted as being the time lag with reference to 0.5--1.5 keV band.

For the bright hard state observation, the model predictions match well with the LAXPC observations for frequencies $\sim 0.1$ and $\sim 10$ Hz, with reduced $\chi^2_{r} = 0.8$ and $1.2$, respectively. For several observations of Cygnus X--1 the best-fit parameters of $\delta T_{\rm e}$, $\delta T_{\rm in}/\delta T_{\rm e}$ and  $\tau_D$ range from 0.005--0.03, 0.6--1.8 and 70--300 ms for $\sim 0.1$ Hz while for $\sim 10$ Hz the corresponding ranges are 0.015--0.03, 0.3--0.6 and 2--9 ms \citep{Maq19}. For MAXI J1820+070, the values obtained are $\sim 0.03$, $\sim 1.2$ and $\sim 500$ ms for $0.1$ Hz and $\sim 0.03$, $\sim 0.6$ and $\sim 6$ ms for 10 Hz. The parameter values obtained in this work for \maxij{} are in the same range as the above (Table \ref{modelfit}). Extending the time lag model predictions to lower energies shows a reasonable match with the {\it NICER} results for 0.1 and 10 Hz (Right top and bottom panels of Figure  \ref{rms_lag_as5_n5_n6_3freq}). For the rms variation, {\it NICER} results are above the predictions for 0.1 Hz while for 10 Hz there is an under prediction. Note that the model predicts a turnover at $\sim 2$ keV, and hence {\it NICER} data in principle should be able verify the prediction. However, we caution against over interpretation since the {\it NICER}  and LAXPC results are not strictly simultaneous and more importantly, the model is only applicable to the Comptonized component and at these low energies the disk emission could be important.

For the results corresponding to $\sim 1$ Hz, the model fit is not acceptable with  $\chi^2_{r} = 2.9$, although the parameters obtained are similar to what has been estimated for Cygnus X--1. An inspection of the LAXPC results show that the energy-dependent time lag (middle right panel of Figure \ref{rms_lag_as5_n5_n6_3freq}) is well constrained with relatively smaller error bars and has a monotonic behaviour with an inflection point around 20 keV. This complexity is not captured by the model which predicts only smooth behaviour. This behaviour could be generic and has been brought out due to better statistics at $\sim 1$ Hz or it could be due to the presence of the QPO at $\sim 0.91$ Hz which may have a different temporal behaviour than the broadband noise. Extension of the model prediction of the time lag to lower energies shows a clear mismatch with {\it NICER} results and the predicted rms is lower than what is observed (middle panel of Figure \ref{rms_lag_as5_n5_n6_3freq}). Interestingly, there is a turnover at $\sim 2$ keV for the {\it NICER} results as predicted by the model but at a higher rms level. Again for reasons mentioned above, we caution against over interpreting  comparison of the {\it NICER} results with the extension of the model to lower energies.


For the faint hard state the statistics are not good enough to perform detailed modelling of the energy-dependent variability. Nevertheless, we modelled the fractional rms and time lag for the LAXPC data for frequency range 0.8-1.2 Hz. The best-fit models are shown as solid lines in Figure \ref{rms_lag_as4} and parameters are listed in Table \ref{modelfit}. It is interesting to note that the parameter values are similar to those obtained for the bright hard state data for the same frequency range, except that the time delay $\tau_D$ is negative. A negative $\tau_D$ means that the soft photon source variation $\delta T_{in}$ occurred after the variation in the coronal temperature $\delta T_e$ which implies that the variability originates in the corona and propagates outward in contrast to the more standard propagating model interpretation used for the bright hard state in this work, as well as for Cygnus X--1 \citep{Maq19} and MAXI J1820+070 \citep{Mud20}. Indeed, an identical outward propagating interpretation was invoked by \citet{Jit19} to explain the soft lags observed for the QPO in Swift J1658.2-4242 in its hard intermediate state. However, it should be emphasised that the significance of the time lag measurement in the LAXPC data of the faint state is not as high as that it is for the high flux hard state data.

\begin{table}
\setlength{\tabcolsep}{3.0pt}
	\caption{Best-fit Parameters from the Stochastic Propagation Model. (1) Spectral state (F and B represent the faint and bright hard state, respectively); (2) Frequency range in Hz; (3) ratio of the variation in the inner disk temperature to the variation of electron temperature; (4) time lag in ms; (5) variation in the electron temperature; (6) $\chi^2$ statistics and degrees of freedom.}
 	\begin{tabular}{@{}cccccc@{}}
	\hline
	\hline
State & Freq & $\delta T_{\rm in}/\delta T_{\rm e}$  & $\tau_{D}$ & $\delta T_{\rm e}$ & $\rm \chi^2_{r}/ d.o.f$ \\
\hline

F & 0.8-1.2     & $0.61^{+0.97}_{-0.32}$ & $-93.46^{+61.75}_{-56.98}$ & $0.025^{+0.013}_{-0.013}$ & 0.6/7 \\
B & 0.08-0.12   & $1.09^{+0.10}_{-0.09}$ & $342.26^{+60.77}_{-60.51}$ & $0.011^{+0.001}_{-0.001}$ & 0.8/17 \\
  & 0.8-1.2     & $0.86$ & $71.86$ & $0.024$ & {\bf 2.9}/17 \\
  & 8-12    & $0.52^{+0.05}_{-0.05}$ & $9.07^{+1.21}_{-1.23}$ & $0.021^{+0.001}_{-0.001}$ & 1.2/17 \\

\hline
\end{tabular} 
\label{modelfit}
\end{table}

\section{Summary and Discussion}
\label{sec:discu}

In this work, we have presented the broadband spectral-timing analysis of the new black hole binary candidate \maxij{} using simultaneous {\it AstroSat} and {\it NICER} observations. The main results of the study are summarized below.

\begin{itemize}

\item The broadband spectral analysis (in the 0.8--40 keV energy) of the five {\it AstroSat} observations identified the source to be in the soft spectral state for the first three observations and in the hard state for the last two observations. The two hard state observations differ significantly in flux and hence were named as faint and bright hard states. The soft state spectra are disk emission dominated with a high energy power-law index $\sim 2$, while the hard state spectra are dominated by Comptonized emission with power-law index $\sim 1.55$.

\item In one of the soft state {\it AstroSat} observations a weak ($< 2\%$ rms), broad ($Q < 2$) QPO is detected with a centroid frequency of $\sim 6.9$ Hz. This low frequency QPO most likely belongs to the class of type-A QPOs \citep{Cas04, Bel11, Mot16}. Another QPO with a sub-harmonic feature, was detected in the bright hard state observation, with centroid frequency $\sim 0.9$ Hz, variability amplitude $\sim 6$\% and a Q factor of $\sim 4.7$. From these properties, we identify this QPO as a type-C QPO \citep{Cas04, Cas05, Mot15}.

\item For the first time, we estimated the energy-dependent fractional rms and time lag of \maxij{} in the 0.5--80 keV energy band using the {\it NICER}/XTI and {\it AstroSat}/LAXPC instruments for a range of frequencies and for the QPOs.

\item In the soft state observations, the power density spectra (computed in the 0.01--30 Hz frequency range) for the 0.5--12 keV {\it NICER} band is significantly lower by at least a factor of $\sim 5$ from that of the 3--80 keV {\it AstroSat} LAXPC. This is further illustrated by computing the energy dependence of the fractional rms at different frequencies, which show an increasing trend with energy.
Based on the spectral fitting, this implies that the variability in the soft state is dominated by the hard X-ray Comptonized component and the disk emission is significantly less variable.

\item For the bright hard state observation, fitting the power density spectra with Lorentzian functions, revealed that the fractional rms of the Lorentzian representing the QPO at $\sim 0.9$ Hz is nearly energy independent at $\sim 5$\%, while a decreasing trend with energy is seen for the sub-harmonic. However, when the fractional rms is estimated from the PDS directly in a broad frequency range of 0.8--1.2 Hz a clear decrease with energy is detected. Moreover, the fractional rms is different for {\it NICER} and LAXPC data even in the same energy band. This implies that the temporal features around $\sim 1$ Hz is complex with the presence of a weak QPO along with broad band noise. The inconsistency between the {\it NICER} and LAXPC fractional rms estimation can be due to variation of the source's temporal property at $\sim 1$ Hz during the observation. Strictly simultaneous PDS generated from {\it NICER} and LAXPC show similar shapes although the statistics are low due to the smaller exposure time. The fractional rms in the frequency ranges of 0.08--0.12 and 8.0--12.0 Hz decreased and moderately increased with energy, respectively and the {\it NICER} and LAXPC data points were consistent with each other in the common energy range.

\item For the bright hard state observation, hard time lags (i.e. time lags increasing with energy) are clearly detected at $\sim 0.1$, $1$ and $10$ Hz in the unprecedented energy range of 0.5--80 keV. The time lag between 60 keV and 1 keV photons varies with frequency such that it is $\sim 400$, $\sim 80$ and $\sim 8$ milli-seconds at $\sim 0.1$, $1$ and $10$ Hz, respectively. At $\sim 1$ Hz the time lag shows monotonic behaviour with an inflection point around 20 keV which may be related to the complexity of having a weak QPO along with broad band noise in that frequency range.

\item For the faint hard state observation, the PDS of the {\it NICER} and LAXPC observations were found to be similar with enhanced variability in the higher energy LAXPC band for frequencies $> 2$ Hz. Soft time lag (i.e. time lag decreasing with energy) was detected in the LAXPC band at $\sim 1$ Hz.

\item We fitted the energy-dependent fractional rms and time lags using a simple single-zone stochastic propagation model \citep{Maq19}. The model is parametrized by variation in the input seed photon temperature ($\delta T_{in}$), the coronal electron temperature ($\delta T_{e}$) and the time lag between them ($\tau_D$). It requires the time averaged spectral parameters of the Comptonization component and is valid only when the time averaged spectrum is dominated by the Comptonization component. The model describes the LAXPC 3--80 keV data well for the bright hard state for frequencies $\sim 0.1$ and $\sim 10$ Hz, but fails to fit the monotonic nature of the time lag at $\sim 1$ Hz. The parameters obtained are similar to the ones obtained from fitting the energy-dependent fractional rms and time lags of Cygnus X--1 and MAXI J1820+070 \citep{Maq19, Mud20}. For the faint hard state the soft lags require that $\tau_D$ to be negative i.e. the coronal temperature varies before the seed photon one.

\item Extending the single-zone stochastic model fitted to LAXPC data to lower energies, we find that the predicted rms and time lag are qualitatively similar but quantitatively different from {\it NICER} results, especially at $\sim 1$ Hz. This discrepancy could be because the {\it NICER} and LAXPC data are not strictly simultaneous and/or the model does not take into account disk emission which
contributes in the low energy band.

\end{itemize}

The QPO detections presented here are consistent with previous studies of the source with {\it NICER} observations \citep{Zha20}. However, {\it NICER} also detected a $\sim 7$ Hz type-A QPO in four data segments five days after the AS3 observation \citep{Zha20}. In addition, a strong type-B QPO at $\sim 4.5$ Hz in the SIMS was a detected in a set of {\it NICER} observations \citep{Bel20}. Unfortunately, {\it AstroSat} did not observe the source during these times.

It is known that the variability is weak in the soft state of BHXRBs \citep[e.g.][]{Mot16} and that the disk emission is significantly less variable than the Comptonized one. However, with LAXPC and {\it NICER} observations we could measure the variability across a wide range of energies and hence confirm that the emission below 4 keV is significantly less variable than that for higher energies. Moreover, spectral analysis revealed that indeed the disk emission contributes significantly below 4 keV.

Rapid repeated flux variations have been detected in a handful of black hole X-ray transients during their outburst, which are generally referred to as flip flops \citep{Miy91, Tak97, Sri12, Bog20, Bui21}. These flip flops are characterised by an abrupt change in flux with transition time scales ranging from few tens to more than 1 ksec \citep{Miy91, Tak97, Hom01, Hom05}. They occur in the intermediate state and exhibit rapid transitions between different types of QPOs. On some occasions, significant changes in spectral parameters have been observed \citep{Bog20}, although for others the spectral parameters remain unchanged \citep{Miy91}. Here, for MAXI J1348--630, we have seen a reminiscent of flip-flops in one of the {\it AstroSat} observations (AS2; see middle panel of Figure \ref{lc_soft}). It is interesting to note that this rare phenomenon is observed here in the soft state and we do not detect QPOs in the bright and dim phases of flip flops in contrast to previous studies. However, there is evidence for a broad feature in the PDS extracted from the high-intensity level (see middle panel of Figure \ref{pds_first3}). Moreover, for MAXI J1348--630, we do observe changes in the spectral parameters similar to previously detected flip flops. In particular the photon index, scattering fraction and the inner disk temperature change during the flip flop event.

{\it AstroSat} observations provide simultaneous spectral coverage from 0.8--40 keV along with energy-dependent fractional rms and time lag in the 3--80 keV band, for a range of Fourier frequencies.
This has allowed to test a physical (albeit simple) fluctuation model and to obtain physical parameters \citep{Maq19}. For the hard state of Cygnus X--1, the physical parameters consisting of variation of inner disk temperature, coronal temperature and the time-delay between them varied for different observations \citep{Maq19}. The same model could also explain the timing features of MAXI J1820+070 in the hard state \citep{Mud20}. While the above analysis were undertaken for the continuum variability, the model has also been applied to a QPO observed in Swift J1658.2--4242 \citep{Jit19}, where soft instead of hard time lags were seen. This implied that for the QPO in Swift J1658.2--4242, the coronal temperature varied before the inner disk one, which in turn suggests that the variability is propagating in the outward direction. Here, we show that for the bright hard state of \maxij{}, the model fits the data for $\sim 0.1$ and $\sim 10$ Hz, but deviations are seen for $\sim 1$ Hz. The parameter values obtained are similar to the ones obtained for Cygnus X--1 and MAXI J1820+070. For the faint hard state data, soft lags are observed and hence model fitting reveals coronal temperature variation earlier than the inner disk one, similar to the results obtained for the QPO in Swift J1658.2--4242.

{\it NICER} data allows for verifying the model predictions at energies lower than 4 keV. We find that while there is qualitative similarity between the model predictions and {\it NICER} measurements of the fractional rms and time lags, there are quantitative differences. It is important to note that the model used in this work is only valid for the thermal Comptonized component. The presence of disk emission at these energies would need to be considered. A more sophisticated model incorporating the disk emission has been formulated and applied to the energy-dependent variability of a QPO of GRS 1915+105 \citep{Gar20}. However, since the source's rapid temporal behaviour may vary in time-scales of hours, such models can be applied with confidence only to strictly simultaneous {\it NICER} and
{\it AstroSat} observations.

It should be emphasized that the data used in this work from {\it NICER} and {\it AstroSat} are not from a coordinated observation between the missions. Hence, the strictly simultaneous data from the two mission is sparse which has limited the analysis. This underlines the need for joint coordinated observations between {\it NICER} and {\it AstroSat} in the future, which would be critical for our understanding of the broadband spectral-timing properties of X-ray binaries.

\section*{Acknowledgements}
We thank the anonymous referee for the constructive comments and suggestions that improved this manuscript.
VJ thanks Liang Zhang, Diego Altamirano and Sunil Chandra for the useful discussion related to the {\it NICER} data analysis and SXT pile-up issues. GM acknowledges the support from the China Scholarship Council (CSC), Grant No.~2020GXZ016647. The research is based on the results obtained from the {\it AstroSat} mission of the Indian Space Research Organization (ISRO), archived at the Indian Space Science Data Centre (ISSDC). This work has used the data from the LAXPC and SXT instruments. We thank the LAXPC Payload Operation Center (POC) and the SXT POC at TIFR, Mumbai for providing the data via the ISSDC data archive and the necessary software tools. This research has made use of data and software provided by the High Energy Astrophysics Science Archive Research Center (HEASARC), which is a service of the Astrophysics Science Division at NASA/GSFC.\\

\section*{Data Availability}

The data used in this article are available in the ISRO's Science Data Archive for {\it AstroSat} Mission (\url{https://astrobrowse.issdc.gov.in/astro_archive/archive/Home.jsp}) and HEASARC database ({\url{https://heasarc.gsfc.nasa.gov}). The source code for the model used in the paper can be shared on reasonable request to the corresponding author, V. Jithesh (email: vjithesh@iucaa.in or jitheshthejus@gmail.com).









\bsp	
\label{lastpage}
\end{document}